\documentclass[10pt,a4paper]{article}

\usepackage{amsmath}
\usepackage{amssymb}
\usepackage{slashed}
\usepackage[dvips]{graphics}
\usepackage[dvips]{color}
\usepackage{graphicx}
\usepackage{multicol}

\newcommand{\be}{\begin{equation}} \newcommand{\ee}{\end{equation}}
\newcommand{\ba}{\begin{eqnarray}} \newcommand{\ea}{\end{eqnarray}}
\newcommand{\bas}{\begin{eqnarray*}} \newcommand{\eas}{\end{eqnarray*}}
\newcommand{\nn}{\nonumber} \newcommand{\mc}{\mathcal}

\makeatletter

 \def\section{\@startsection {section}{1}{\z@}{24pt plus 2pt minus 2pt}
{12pt plus 2pt minus 2pt}{\large\bf}}

\newenvironment{inlinetable}{%
\def\@captype{table}%
\noindent\begin{minipage}{0.999\linewidth}\begin{center}}
{\end{center}\end{minipage}\smallskip}

\newenvironment{inlinefigure}{%
\def\@captype{figure}%
\noindent\begin{minipage}{0.999\linewidth}\begin{center}}
{\end{center}\end{minipage}\smallskip} \makeatother

\begin{document}

%%%%%%%%%%%%%%%%%%%%%%%%%%%%%%%%%%%%%%%%%%%%%%%%%%%%%%%%%%%%%%%%%%%%%%%%%%%%%%%%
%
%        Place for title and abstract
%
%%%%%%%%%%%%%%%%%%%%%%%%%%%%%%%%%%%%%%%%%%%%%%%%%%%%%%%%%%%%%%%%%%%%%%%%%%%%%%%%
\noindent{\Large\sf\textbf{Two Photon Exchange Contributions to
Elastic $\vec e + p \rightarrow e + \vec p$ ~Process in a\\\\
Nonlocal Field Formalism}}

\vspace{5mm} \noindent Pankaj Jain, Satish D. Joglekar and
Subhadip Mitra

\vspace{5mm} \noindent Department of Physics, IIT Kanpur, Kanpur -
208016, India

\vspace{5mm}
\begin{center}
\begin{minipage}{0.8\linewidth}
{\small {\bf Abstract:} We compute the two photon exchange
contributions to elastic scattering of polarized electrons from
target protons. We use a nonlocal field theory formalism for this
calculation. The formalism maintains gauge invariance and provides
a systematic procedure for making this calculation. The results
depend on one unknown parameter $\bar b$. We compute the two
photon exchange correction to the ratio of electric to magnetic
form factors extracted using the polarization transfer
experiments. The correction is found to be small if $\bar b\sim
1$. However for larger values of $\bar b>3$, the correction can be
quite significant. The correction to the polarization transfer
results goes in the right direction to explain their difference
with the ratio measured by Rosenbluth separation method. We find
that the difference between the two experimental results can be explained
for a wide range of values of the parameter $\bar b$. We
also find that the corrections due to two photon exchange depend
on the photon longitudinal polarization $\varepsilon$.
Hence we predict an $\varepsilon$ dependence of the form factor
ratio extracted using the polarization transfer
technique. Finally we obtain a limit on $\bar b$ by requiring that the
non-linearity in $\varepsilon$ dependence of the unpolarized reduced
cross section is within experimental errors.}
\end{minipage}\end{center}
%\titlerunning{}
%\maketitle
%%%%%%%%%%%%%%%%%%%%%%%%%%%%%%%%%%%%%%%%%%%%%%%%%%%%%%%%%%%%%%%%%%%%%%%%%%%%%%%%
\vspace{5mm}
\begin{multicols}{2}

\section{Introduction}
\noindent The observed discrepancy \cite{ARZ,PPV} in the proton
electromagnetic form factors in the JLAB polarization transfer
experiments \cite{Jones,Gayou,Punjabi,Polarization} and the SLAC
Rosenbluth separation experiments
\cite{Walker,Andivahis,Rosenbluth} has been studied extensively in
the literature. The two photon exchange contributions are most
likely the source of the difference
\cite{Guichon,Arrington,Tjon,Brodsky,Rekalo,Belushkin1}. In a
recent paper we constructed a nonlocal Lagrangian to model the
electromagnetic interaction of extended objects such as the proton
\cite{Jain}. The model maintains gauge invariance in the presence
of a form factor at the electromagnetic vertex. We truncate the
Lagrangian to include only operators with dimension five or less.
The dimension five operator is necessary if we include the contribution
proportional to
the Pauli form factor $F_2$. This truncation is reliable as long
as the off-shellness of the proton propagator is small compared to
the hadronic momentum scale. The resulting Lagrangian depends on
the two on-shell form factors, $F_1$ and $F_2$, and contains one
unknown parameter which we denote as $\bar b$. We found that for
small values of this parameter $\bar b\sim 1$, the results of the
SLAC Rosenbluth measurement, after the two photon exchange
correction, are in agreement with the JLAB polarization transfer
results. In the present paper we compute the two photon exchange
corrections to the polarization transfer experiment.

In polarization transfer experiments, longitudinally polarized
electrons are scattered from fixed target protons. The tree level
amplitude for the process (See Fig.~\ref{fig:born}) is given by:
\be \mathcal{M}_0 = - \frac{e^2}{q^2}\bar{u}(k')\gamma^\mu
u(k,s_e) \bar{U}(p',s_{p'})\Gamma_\mu U(p),
\label{pol_born_amp}\ee where \be \Gamma_\mu = F_1(q)\gamma_\mu +
\frac{i\kappa_p}{2M_p} F_2(q)\sigma_{\mu\alpha}q^\alpha.\ee Here
$e$ is the charge of proton, $k$, $k'$ are the momenta of the
initial and final electron, $p$, $p'$ are the initial and final
proton momenta, $M_p$ is the proton mass, $\kappa_p = 1.79$ is
proton anomalous magnetic moment and $q = p'-p = k-k'$ is the
momentum transfer.

Using (\ref{pol_born_amp}), one can calculate the tree level
cross-section for the scattering of longitudinally polarized
electrons from protons: \be \frac{d\sigma^{1\gamma}}{d\Omega_e} =
\frac{|\overline{\mc M_0}|^2E_e^{'2}} {64M_p^2\pi^2E_e^2}, \ee
where $|\overline{\mc M_0}|^2$ represents the amplitude squared,
summed over final electron spin and averaged over initial proton
spin and is given by: \ba \nn|\overline{\mc M_0}|^2
&\hspace{-4mm}=&\hspace{-3mm} \frac12\sum_{s'_e,s_p}|\mc M_0|^2\\
\nn&\hspace{-4mm}=&\hspace{-3mm}
\frac{e^4}{8q^4}\,tr\Big[\big(\slashed{k}+m_e\big)\big(1+\frac{h}{m}\gamma_5\slashed{k}\big)\gamma^\nu\big(\slashed{k}'+m_e)\gamma^\mu\big)\Big]\\
&&\hspace{-3mm}\nn\times\,
tr\Big[\big(\slashed{p}+M_p\big)\gamma_0\Gamma^\dag_\nu\gamma_0\big(\slashed{p}'+M_p\big)\big(1+\gamma_5\slashed{s}_{p'}\big)\Gamma_\mu\Big].\\\ea

\begin{inlinefigure}
\begin{center}
\hspace{8mm}
\includegraphics[scale=0.65]{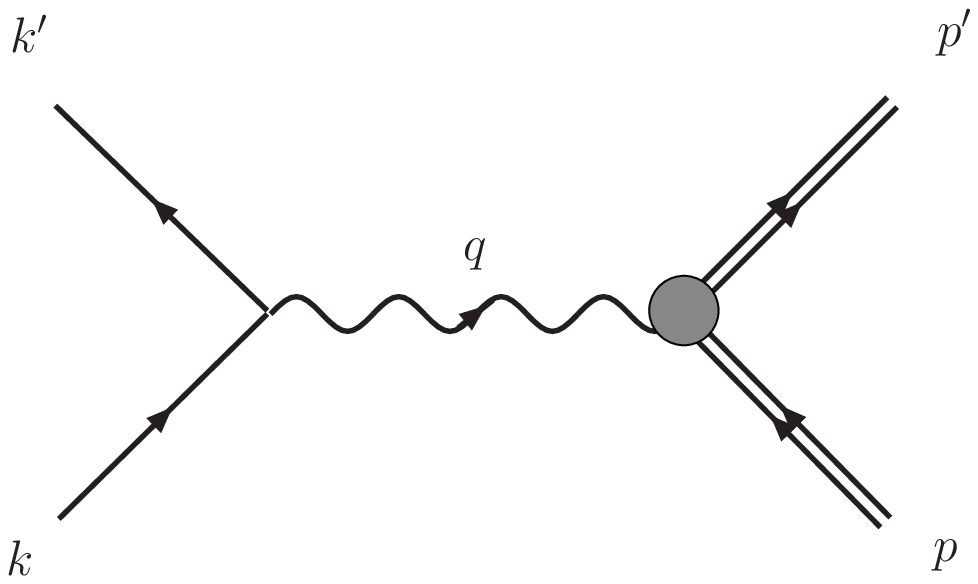}
\label{fig:born} \caption{\small The one photon exchange diagram
contributing to the elastic electron proton scattering. Here $k$,
$k'$ refer to the initial and final electron momenta  and $p$,
$p'$ to the initial and final proton momenta respectively. The symbol $q =
k-k' = p'-p$ denotes the momentum exchanged.\vspace{5mm}}
\end{center}
\end{inlinefigure}
Here $s^\mu_{p'}$ is the spin four-vector of the final proton. In general
the spin
four-vector $s$ for a particle with mass $m$ and momentum
$p=(E;\,\vec p)$ can be written in terms of the unit three-vector
$\hat n$, specifying the spin direction in the rest frame of the
particle, by \be s^\mu =\left(\frac{\hat n\cdot\vec p}{m}; \;\hat
n + \vec p\,\frac{\hat n\cdot\vec p}{m(m+E)}\right).\ee
For the
incident electron, energy $E_e$ is much greater than its mass $m_e$, and we have
approximated the spin four vector,
 $ s^\mu_e \approx h{k^\mu}/{m_e},$
where $h = \hat n_e\cdot\hat k$.

Let the scattering plane be the $X-Z$ plane where the momentum of
the recoiled proton $\vec p\,'$ defines the $Z$ axis, i.e., $\hat z
= \hat p'$. The $Y$ axis is defined as $\hat y = \hat k \times
\hat k'$. This also defines the $X$ axis as $\hat x = \hat y
\times \hat z$. With this choice of coordinate axes, the tree
level cross section can be written as: \ba
\nn\hspace{-4mm}\frac{d\sigma^{1\gamma}}{d\Omega_e}&\hspace{-2mm}=&\hspace{-3mm}
\left(\frac{\alpha^2E'_e\cos^2\frac{\theta_e}{2}}{8E^3_e\sin^4\frac{\theta_e}{2}}\right)
\left(\frac{1}{1+\tau}\right)\\
&\hspace{-4mm}&\hspace{-3mm}\times\Big[I^{1\gamma}_0 +
hI^{1\gamma}_0P^{1\gamma}_{\rm L}(\hat n_{p'}\cdot\hat
z)+hI^{1\gamma}_0P^{1\gamma}_{\rm T}(\hat n_{p'}\cdot\hat x)\Big], \ea
where \bas I^{1\gamma}_0 &=& G^2_E + \frac\tau\varepsilon G^2_{\rm M},\\
I^{1\gamma}_0P^{1\gamma}_{\rm L} &=&
\frac{E_e+E'_e}{M_p}\sqrt{\tau(1+\tau)}G^2_{\rm M}\tan^2\frac{\theta_e}{2},\\
I^{1\gamma}_0P^{1\gamma}_{\rm T}
&=&-2\sqrt{\tau(1+\tau)}G_{\rm M}G_{\rm E}\tan\frac{\theta_e}{2}.\eas Here
$\tau = -q^2/4M_p^2$, $\varepsilon=
1/[1+2(1+\tau)\tan^2(\theta_e/2)]$ is the longitudinal
polarization of the photon and $\theta_e$ is the electron
scattering angle. $I^{1\gamma}_0$ is proportional to the tree
level unpolarized cross-section. $I^{1\gamma}_0P^{1\gamma}_{\rm L}$
($I^{1\gamma}_0P^{1\gamma}_{\rm T}$) is proportional to the term in the
tree level cross-section that corresponds to scattering a
longitudinally polarized electron from an unpolarized proton
producing a longitudinally (transversely) polarized recoiling
proton. Let us define, \ba
\frac{d\sigma^{1\gamma}_0}{d\Omega_e}&=&\left(\frac{\alpha^2E'_e\cos^2\frac{\theta_e}{2}}{8E^3_e\sin^4\frac{\theta_e}{2}}\right)
\frac{I^{1\gamma}_0}{1+\tau},\\
\frac{d(\Delta\sigma^{1\gamma}_{\rm L})}{d\Omega_e}&=&
\left(\frac{\alpha^2E'_e\cos^2\frac{\theta_e}{2}}{8E^3_e\sin^4\frac{\theta_e}{2}}\right)
\frac{I^{1\gamma}_0P^{1\gamma}_{\rm L}}{1+\tau},\\
\frac{d(\Delta\sigma^{1\gamma}_{\rm T})}{d\Omega_e}&=&
\left(\frac{\alpha^2E'_e\cos^2\frac{\theta_e}{2}}{8E^3_e\sin^4\frac{\theta_e}{2}}\right)
\frac{I^{1\gamma}_0P^{1\gamma}_{\rm T}}{1+\tau}.\ea Using these we can
rewrite $d\sigma^{1\gamma}/d\Omega_e$ as \ba
\nn\frac{d\sigma^{1\gamma}}{d\Omega_e} &=&
\frac{d\sigma^{1\gamma}_0}{d\Omega_e}
+h\Big[\frac{d(\Delta\sigma^{1\gamma}_{\rm L})}{d\Omega_e}(\hat n_{p'}\cdot\hat z)\\
&&+\frac{d(\Delta\sigma^{1\gamma}_{\rm T})}{d\Omega_e}(\hat n_{p'}\cdot\hat x)\Big].
\ea
Hence, the
ratio of the form factors is given by: \ba R &=&
\nn\frac{G_{\rm E}}{G_{\rm M}} = -
\frac{P^{1\gamma}_{\rm T}}{P^{1\gamma}_{\rm L}}\frac{E_e+E'_e}{2M_p}\tan\frac{\theta_e}{2}\\
&=&-\frac{d(\Delta\sigma^{1\gamma}_{\rm T})/d\Omega_e}{d(\Delta\sigma^{1\gamma}_{\rm L})/d\Omega_e}\frac{E_e+E'_e}{2M_p}\tan\frac{\theta_e}{2}.\ea

The contribution of the two photon exchange diagrams to the
electron-proton elastic scattering cross section can be written as
\be \frac{d\sigma^{2\gamma}}{d\Omega_e} =
\frac{2Re(\overline{\mathcal{M}^*_0\mathcal{M}_{2\gamma}})E_e^{'2}}
{64M_p^2\pi^2E_e^2}+\mathcal{O}(\alpha^4),
\label{eq:ds2gamma}
\ee where
$\mathcal{M}_{2\gamma}$ is the total amplitude of the two photon
exchange diagrams.  As in the case of tree level process we can
define $d(\Delta\sigma^{2\gamma}_{\rm L})/d\Omega_e$ and
$d(\Delta\sigma^{2\gamma}_{\rm T})/d\Omega_e$ from the terms in
$d\sigma^{2\gamma}/d\Omega_e$ that are proportional to $h(\hat
n_{p'}\cdot\hat z)$ and $h(\hat n_{p'}\cdot\hat x)$ respectively.
Then, \ba \nn\hspace{-4mm}\frac{d\sigma^{2\gamma}}{d\Omega_e}
&\hspace{-2mm}=&
\hspace{-2mm}\frac{d\sigma^{2\gamma}_0}{d\Omega_e}
+\frac{d(\Delta\sigma^{2\gamma}_N)}{d\Omega_e}(\hat n_{p'}\cdot\hat y)\\
&&\hspace{-2mm}+h\Big[\frac{d(\Delta\sigma^{2\gamma}_{\rm L})}{d\Omega_e}(\hat
n_{p'}\cdot\hat z)
+\frac{d(\Delta\sigma^{2\gamma}_{\rm T})}{d\Omega_e}(\hat
n_{p'}\cdot\hat x)\Big] . \ea Here $d\sigma^{2\gamma}_0/d\Omega_e$
represents the terms independent of the spin of the final proton
and $d(\Delta\sigma^{2\gamma}_N)/d\Omega_e$ corresponds to the
normal polarization of the final proton.

Experimentally the polarization components, $P_{\rm L}$ and $P_{\rm T}$
are measured for different $q^2$. From these the ratio of
the form factors is extracted by the following relation:
\be R_{\rm E} =  -
\frac{P_{\rm T}}{P_{\rm L}}\frac{E_e+E'_e}{2M_p}\tan\frac{\theta_e}{2}.
\label{eq:RE}
\ee

\begin{figure*}
\begin{center}
\begin{tabular}{ccc}
\includegraphics[scale=0.58]{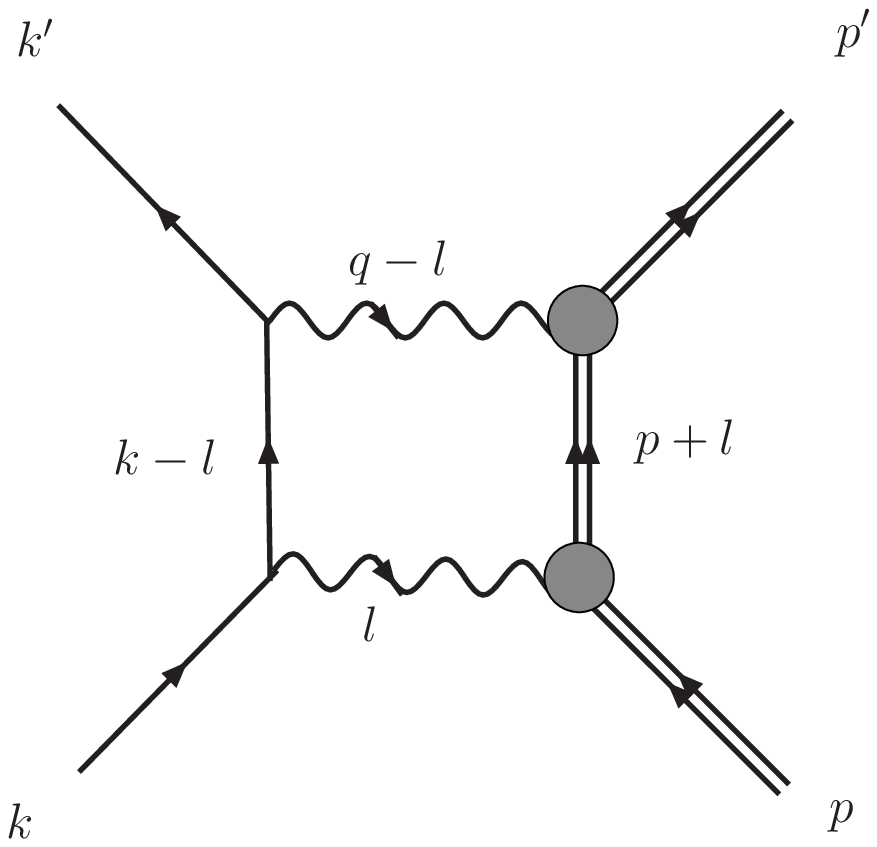}\label{fig:box-a}&\includegraphics[scale=0.58]{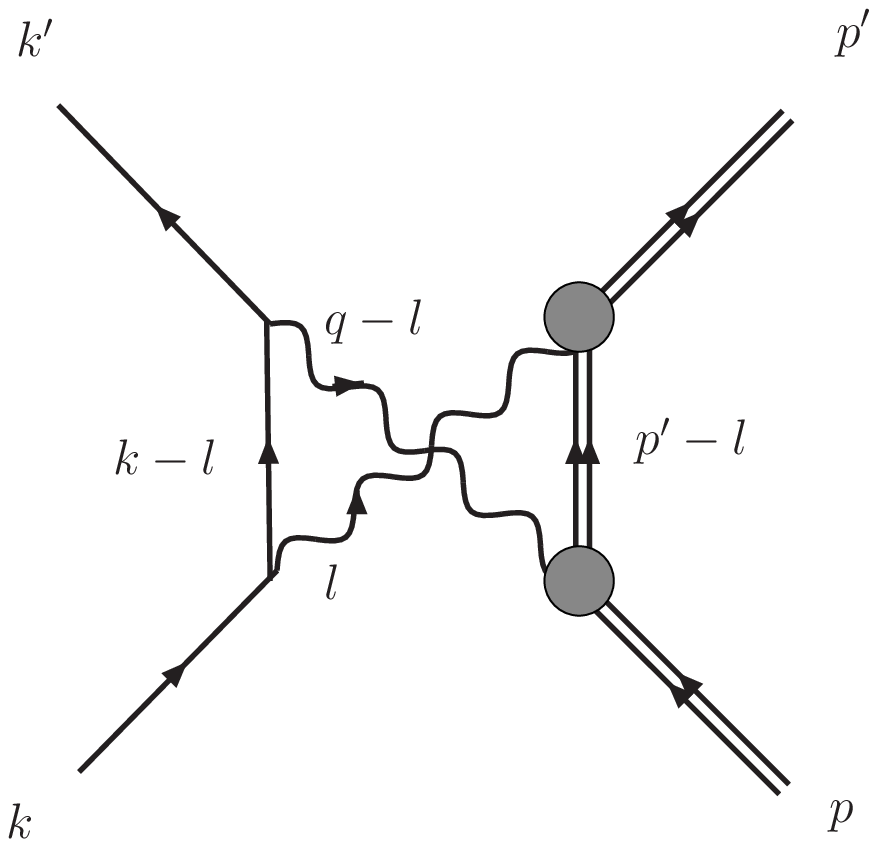}&\includegraphics[scale=0.58]{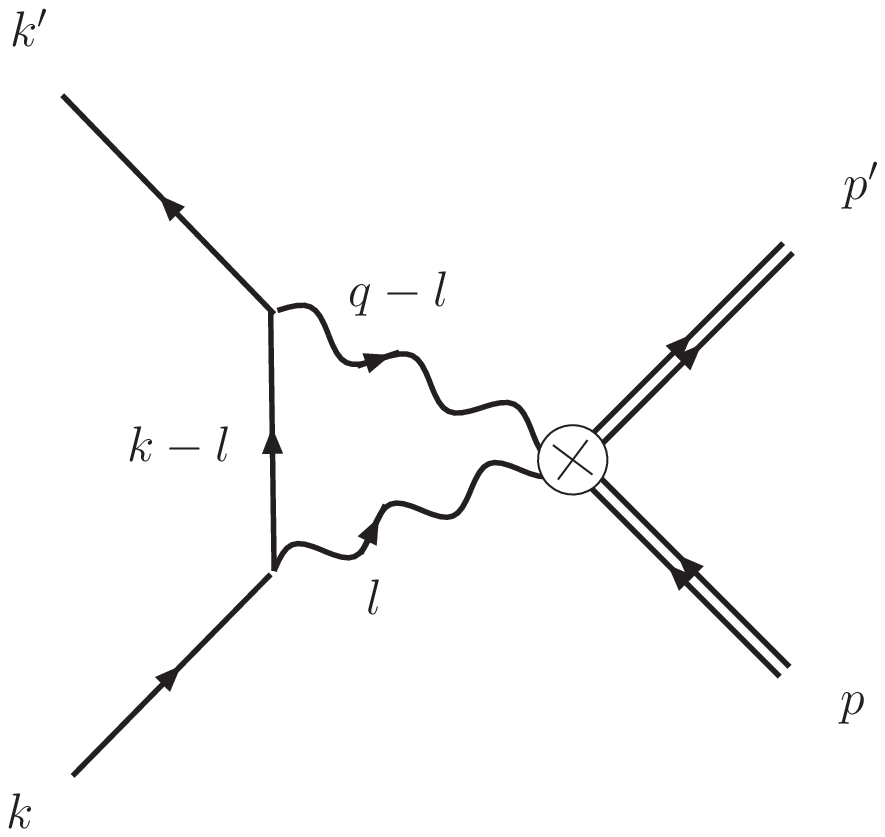}\\
(a)&(b)&(c)\\
\end{tabular}
\caption{\small The two-photon exchange diagrams contributing to
the elastic electron proton scattering: (a) box diagram, (b)
cross-box diagram and (c) diagram 
proportional to $\bar b^2$. Another diagram proportional to 
$\bar b^2$ is obtained by interchanging 
$l\leftrightarrow (q-l$) in (c) } \label{fig:twophoton}
\end{center}
\end{figure*}

\noindent In the tree level approximation $R_{\rm E}$ is identical to $
R$. Let $\widetilde G_{\rm M}$ be the experimentally measured value of $G_{\rm M}$. Then, with the knowledge of $\widetilde G_{\rm M}$ and $R_{\rm E}$ one can define the
following quantities: \ba \frac{d(\Delta\sigma_{\rm L})}{d\Omega_e}&=&
\left(\frac{\alpha^2E'_e\cos^2\frac{\theta_e}{2}}{8E^3_e\sin^4\frac{\theta_e}{2}}\right)
\frac{I_0P_{\rm L}}{1+\tau},\\
\frac{d(\Delta\sigma_{\rm T})}{d\Omega_e}&=&
\left(\frac{\alpha^2E'_e\cos^2\frac{\theta_e}{2}}{8E^3_e\sin^4\frac{\theta_e}{2}}\right)
\frac{I_0P_{\rm T}}{1+\tau}\ea where \ba I_0P_{\rm L} &=&
\frac{E_e+E'_e}{M_p}\sqrt{\tau(1+\tau)}\widetilde G^2_{\rm M}\tan^2\frac{\theta_e}{2},\\
I_0P_{\rm T} &=&-2\sqrt{\tau(1+\tau)}\widetilde G^2_{\rm M}R_{\rm E}\tan\frac{\theta_e}{2}.\ea
From these one obtains the ratio $\bar R$ corrected for the
two-photon exchange processes by the following relation: \ba \nn
\bar R
&=&-\left[\frac{d(\Delta\sigma_{\rm T})/d\Omega_e-d(\Delta\sigma^{2\gamma}_{\rm T})/d\Omega_e}{d(\Delta\sigma_{\rm L})/d\Omega_e-d(\Delta\sigma^{2\gamma}_{\rm L})/d\Omega_e}\right]\\
&&\times\frac{E_e+E'_e}{2M_p}\tan\frac{\theta_e}{2}\\
&=&R_{\rm E}\frac{1-\Delta_{\rm T}}{1-\Delta_{\rm L}}
\label{eq:Rbar}
\ea
where \ba\Delta_{\rm T} &=&
\frac{d(\Delta\sigma^{2\gamma}_{\rm T})/d\Omega_e}{d(\Delta\sigma_{\rm T})/d\Omega_e},\label{eq:delta_t}\\
\Delta_{\rm L}&=&\frac{d(\Delta\sigma^{2\gamma}_{\rm L})/d\Omega_e}{d(\Delta\sigma_{\rm L})/d\Omega_e}\label{eq:delta_l}.
\ea
%\begin{inlinefigure}
%\begin{center}
%\begin{tabular}{c}
%\includegraphics[scale=0.65]{box.eps}\\
%(a)\\
%\includegraphics[scale=0.65]{cbox.eps}\\
%(b)\\
%\end{tabular}
%\caption{\small The two-photon exchange diagrams contributing to
%the elastic electron proton scattering: (a) box diagram and (b)
%cross-box diagram.} \label{fig:twophoton}
%\end{center}
%\end{inlinefigure}
%\begin{inlinefigure}
%\begin{center}
%\includegraphics[scale=0.6]{bpm.eps}
%\caption{\small The two-photon exchange diagrams (obtained by
%$q\leftrightarrow q-l$) proportional to $\bar b^2$.}
%\label{fig:twophotonbbar}
%\end{center}
%\end{inlinefigure}

\begin{figure*}
\begin{center}
\begin{tabular}{cc}
\hspace{-10mm}\includegraphics[scale=0.35,angle=270]{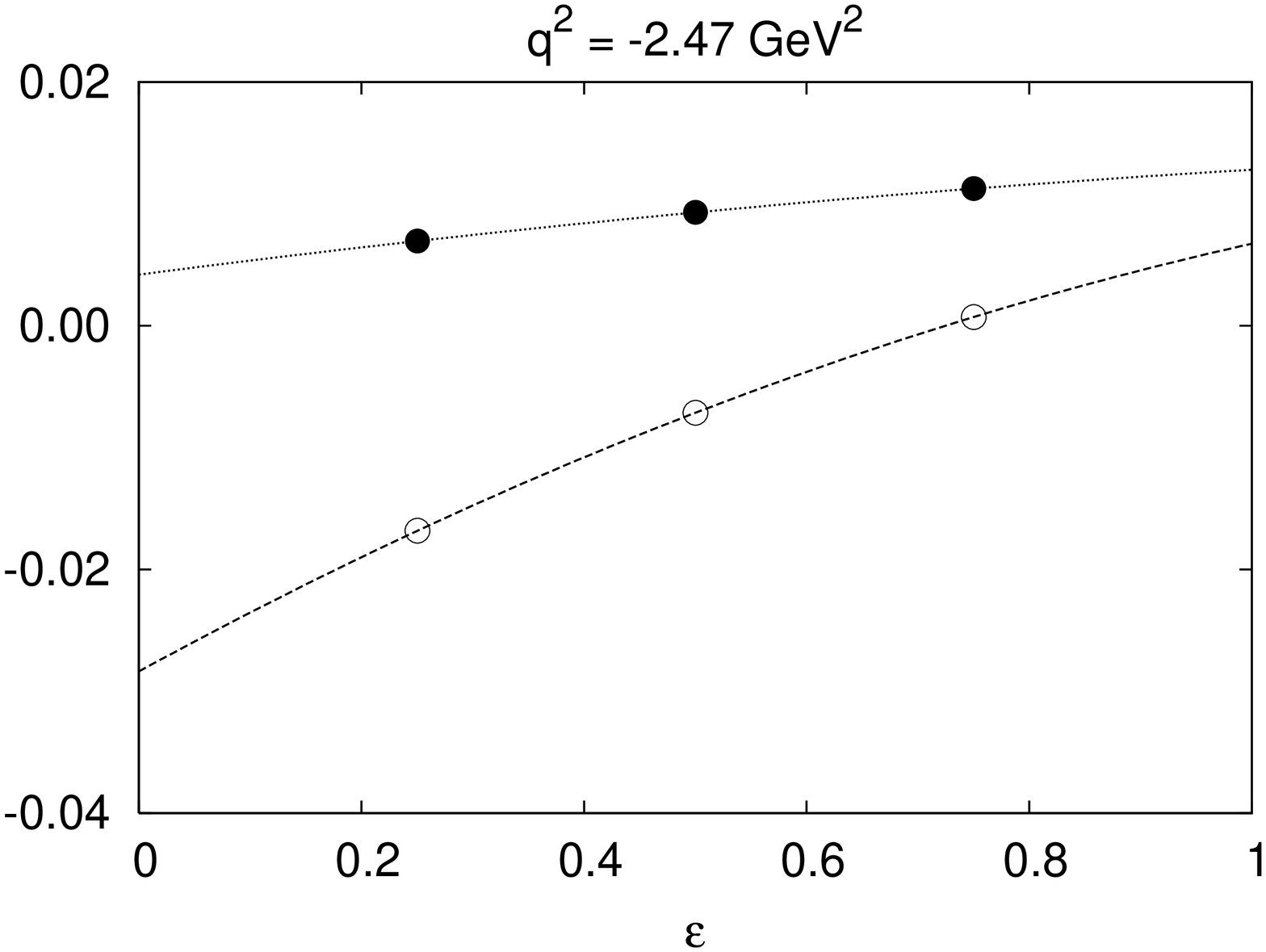}&
\hspace{-10mm}\includegraphics[scale=0.35,angle=270]{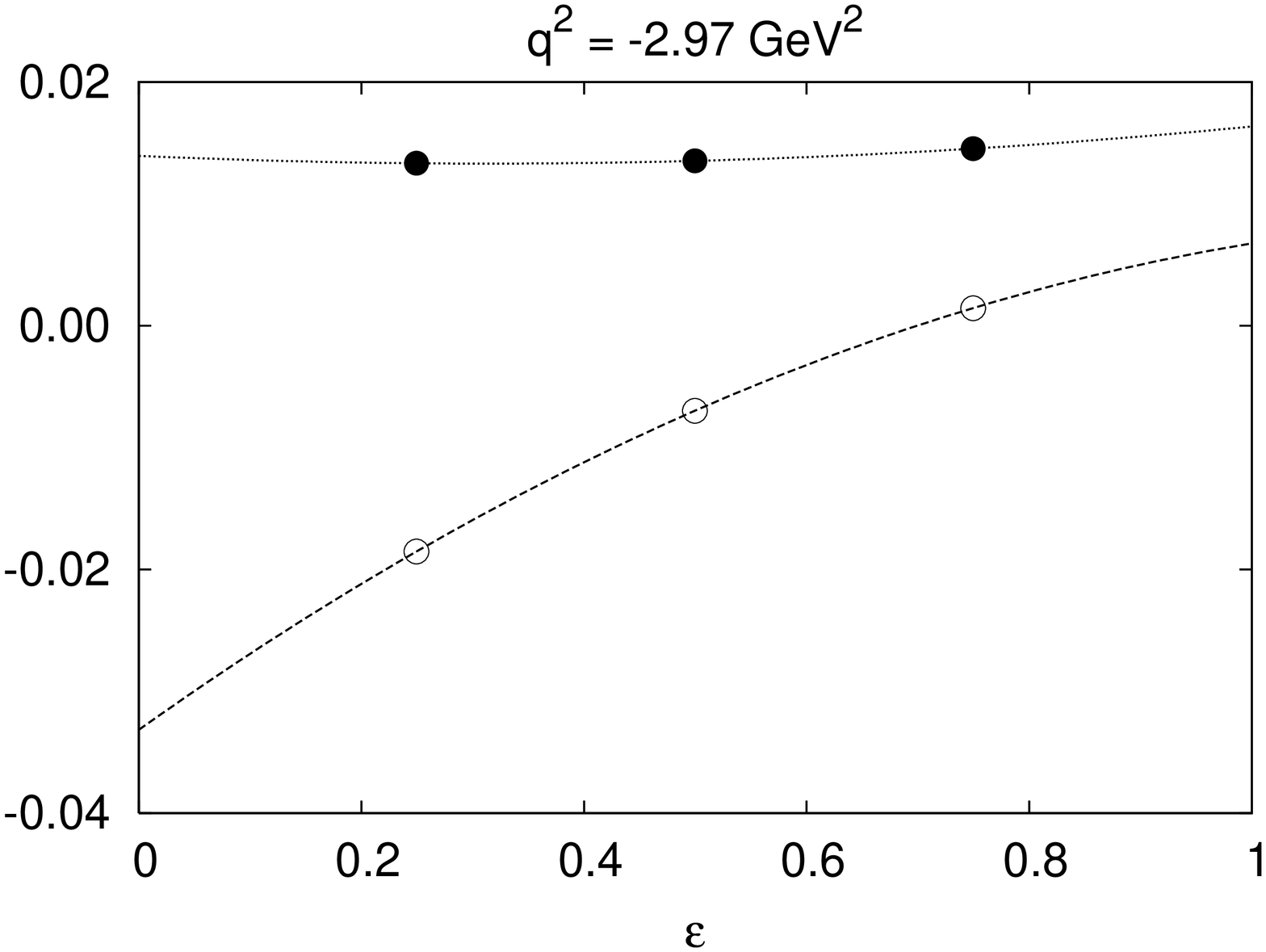}\\
(a)&(b)\\
\hspace{-10mm}\includegraphics[scale=0.35,angle=270]{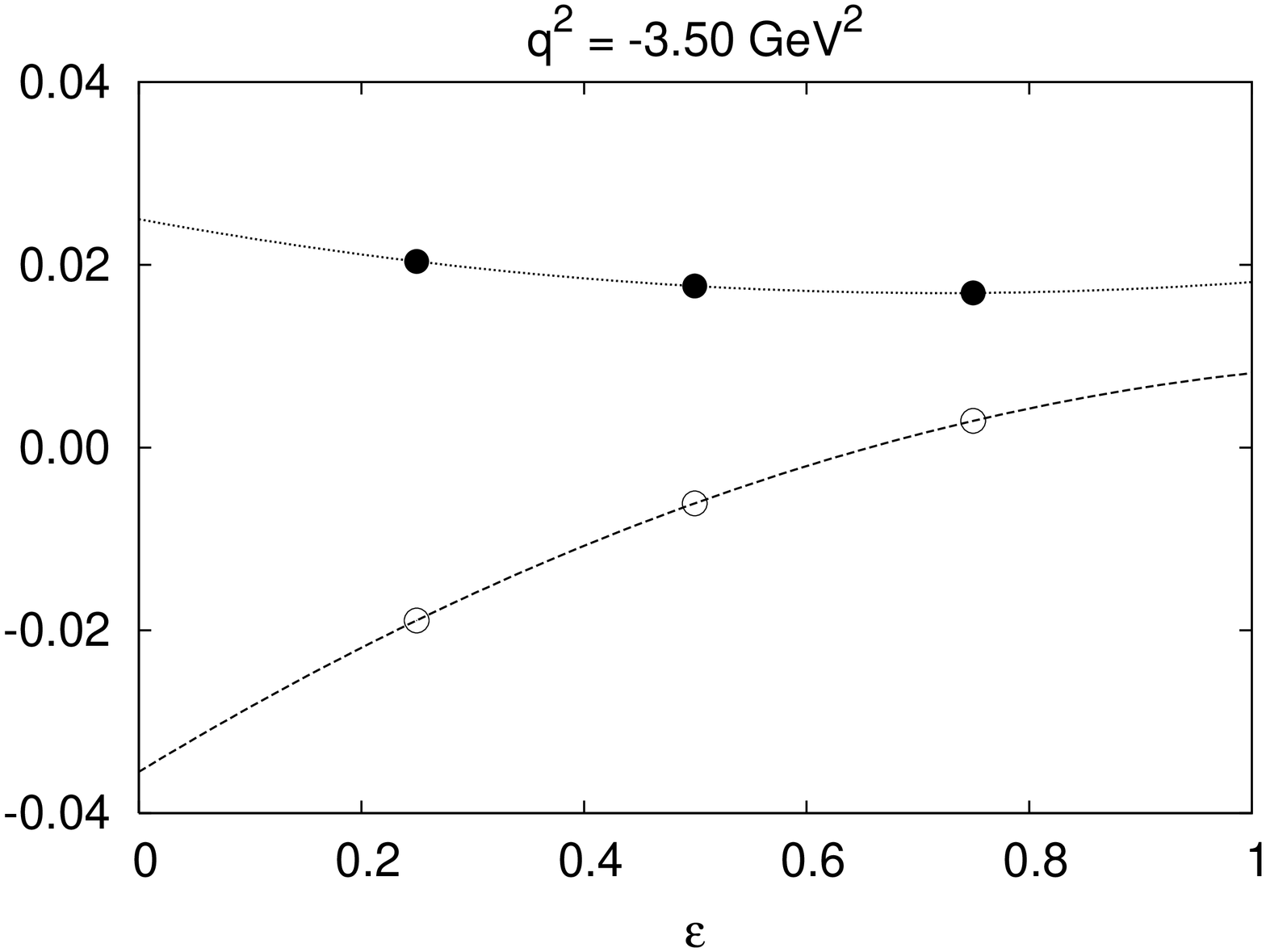}&
\hspace{-10mm}\includegraphics[scale=0.35,angle=270]{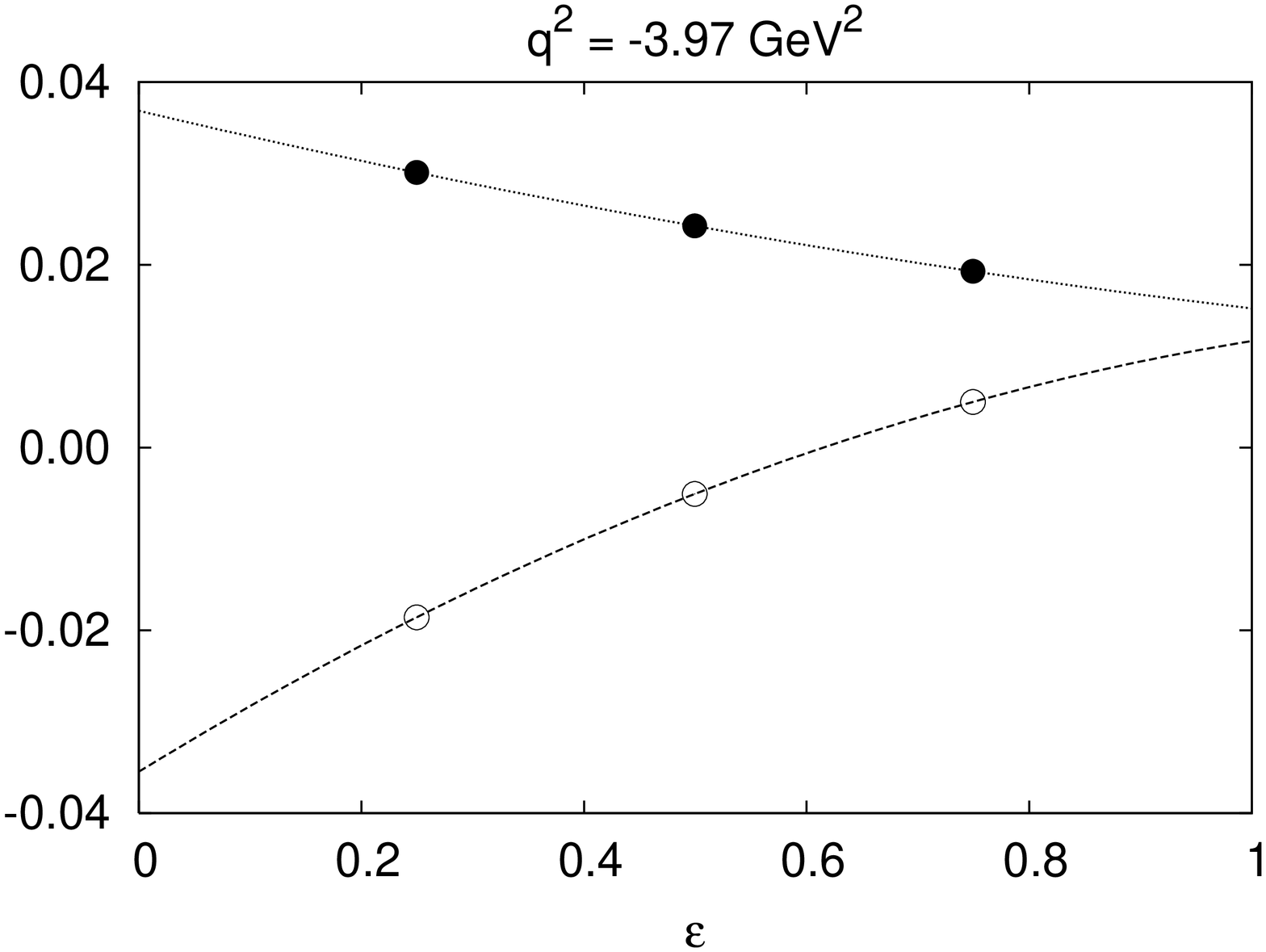}\\
(c)&(d)\\
\hspace{-10mm}\includegraphics[scale=0.35,angle=270]{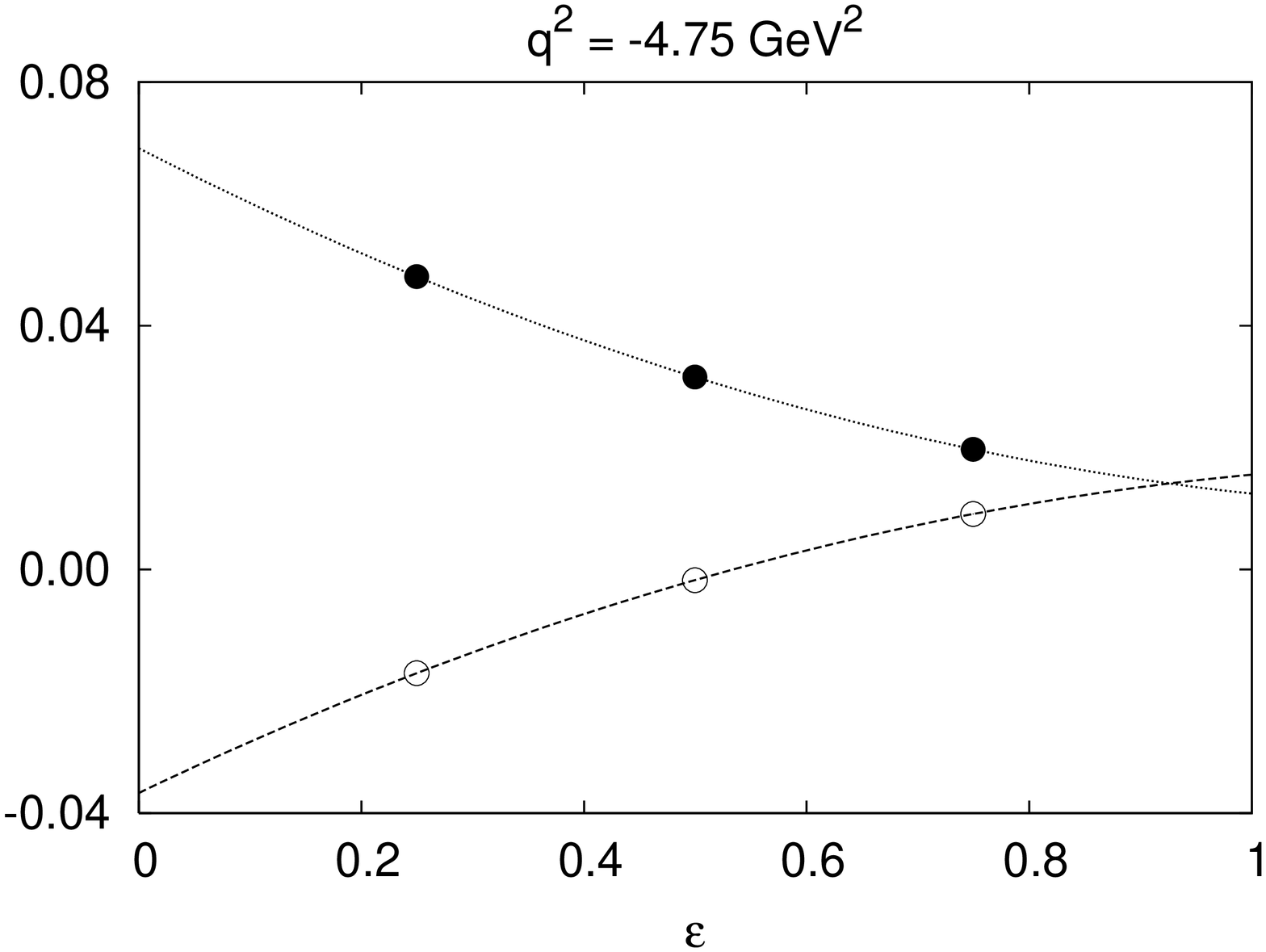}&
\hspace{-10mm}\includegraphics[scale=0.35,angle=270]{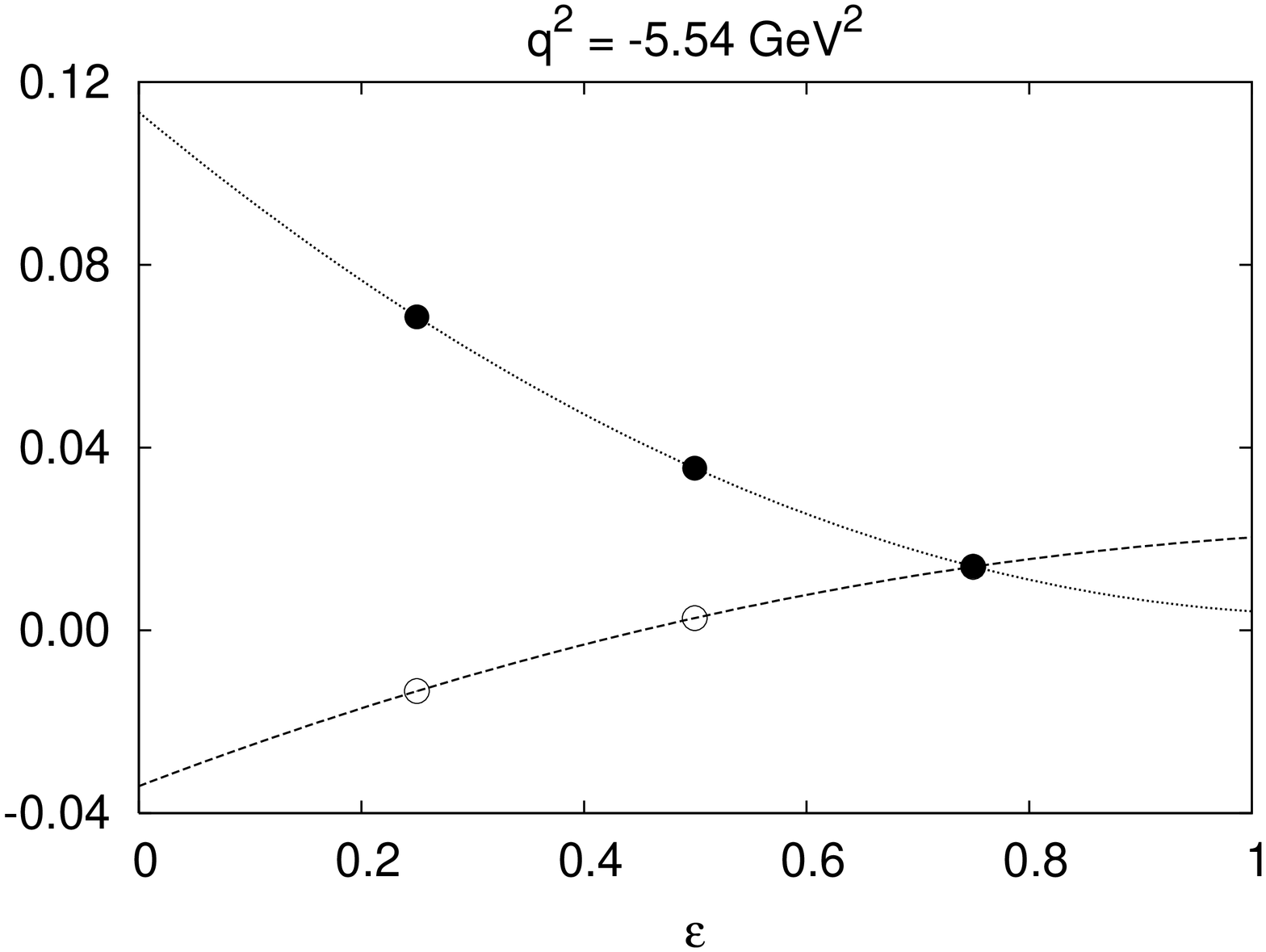}\\
(e)&(f)\\
\end{tabular}
\caption{\small Contribution of box and cross-box diagrams to
$\Delta_{\rm L}$ (unfilled circles) and $\Delta_{\rm T}$ (filled circles) for
different $q^2$ - (a) $q^2 = -2.47 GeV^2$, (b) $q^2 = -2.97
GeV^2$, (c) $q^2 = -3.50 GeV^2$, (d) $q^2 = -3.97 GeV^2$, (e) $q^2
= -4.75 GeV^2$, (f) $q^2 = -5.54 GeV^2$. $\Delta_{\rm T}$ is defined in
(\ref{eq:delta_t}) and $\Delta_{\rm L}$ is defined in
(\ref{eq:delta_l}).} \label{fig:result_bcb}
\end{center}
\end{figure*}
\begin{figure*}
\begin{center}
\begin{tabular}{cc}
\hspace{-11mm}\includegraphics[scale=0.35,angle=270]{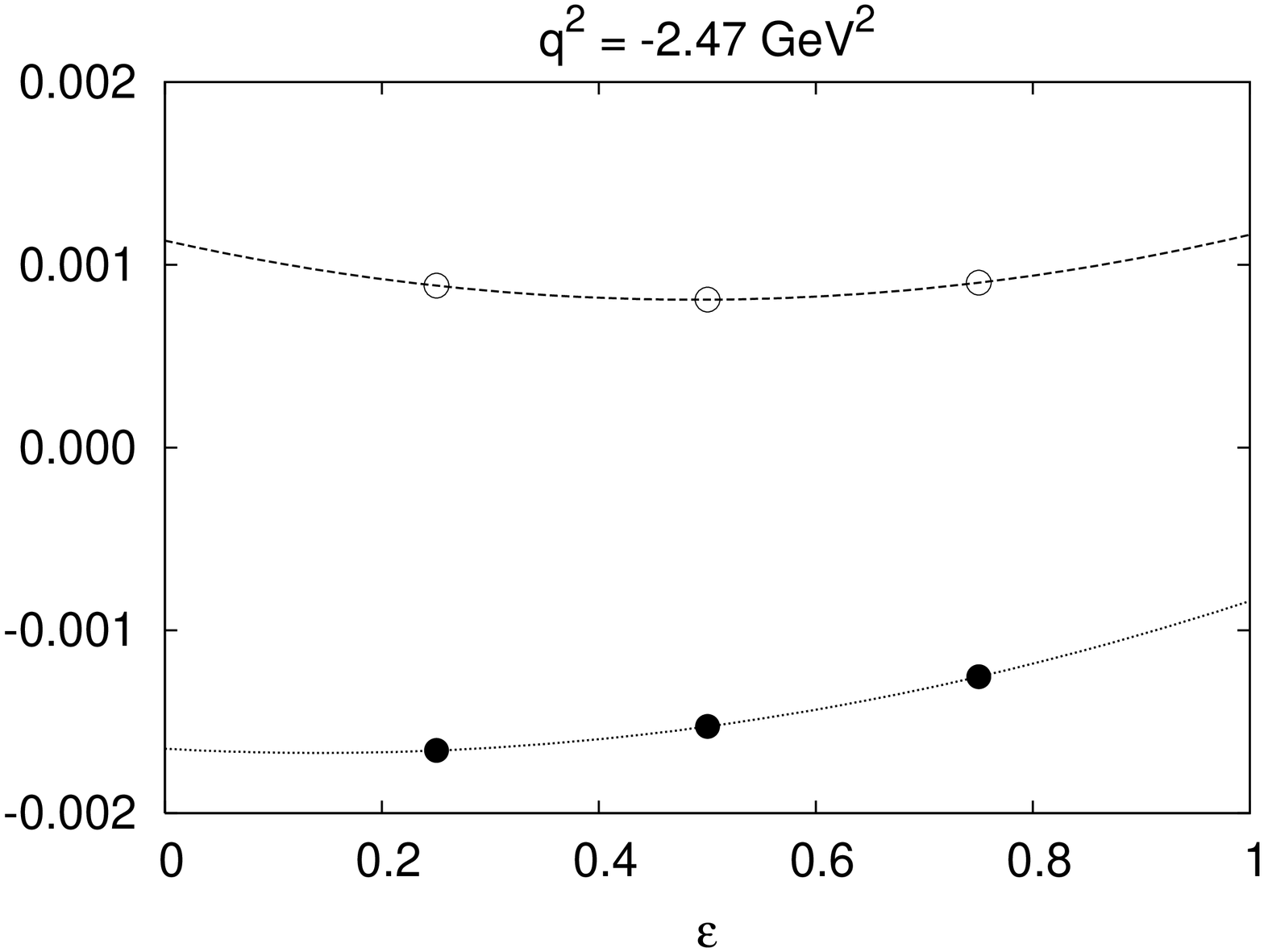}&
\hspace{-11mm}\includegraphics[scale=0.35,angle=270]{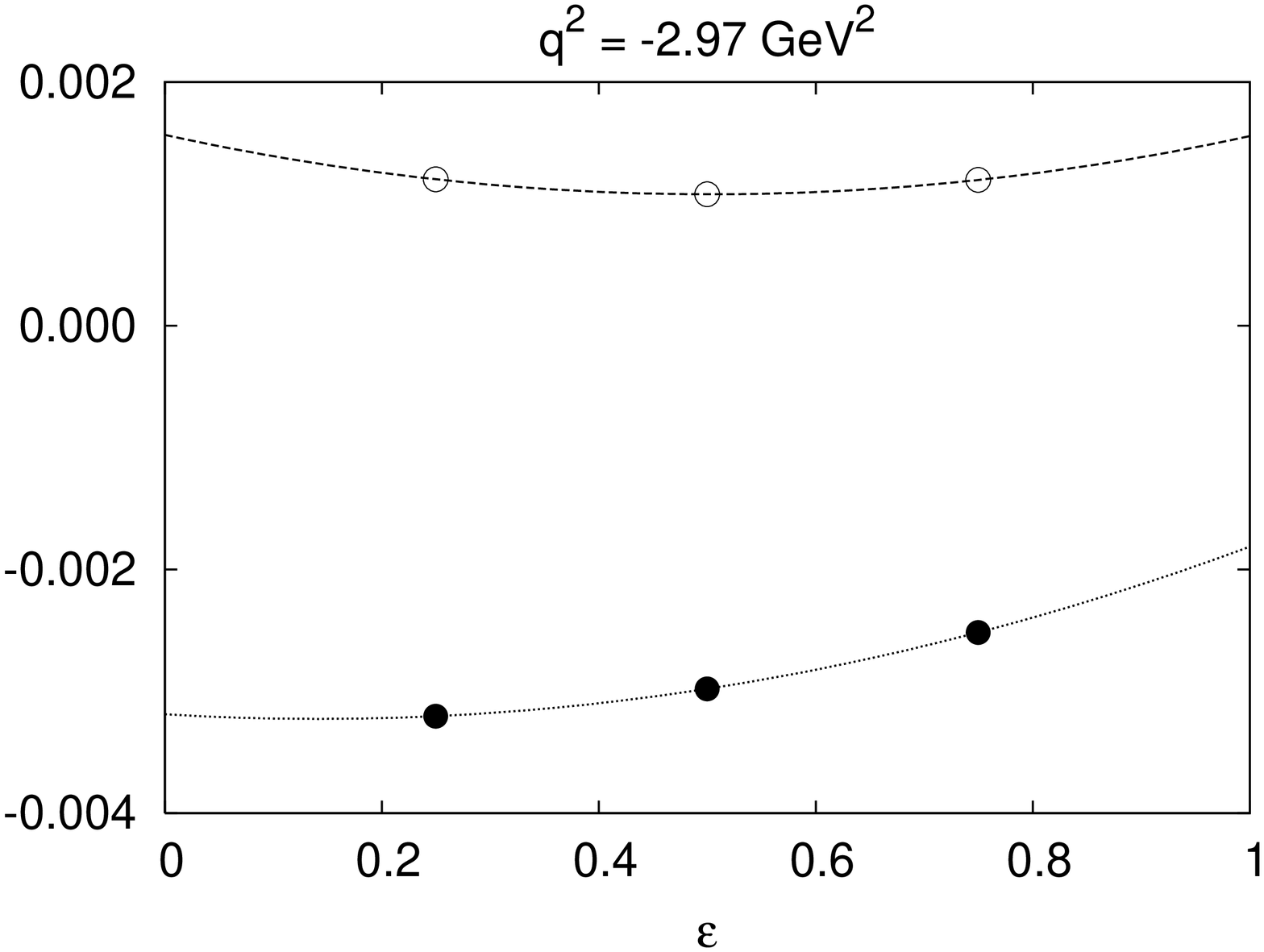}\\
(a)&(b)\\
\hspace{-11mm}\includegraphics[scale=0.35,angle=270]{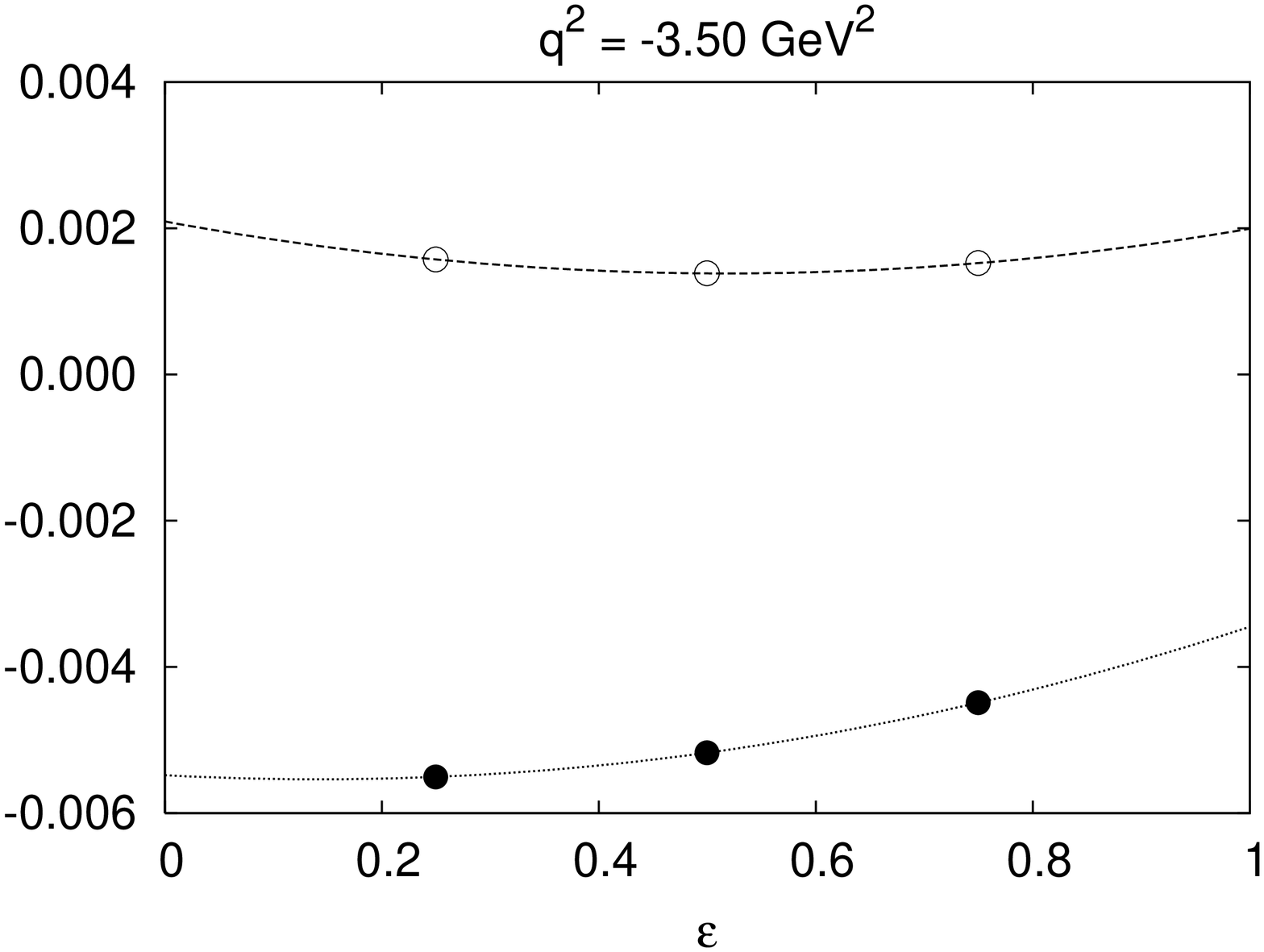}&
\hspace{-11mm}\includegraphics[scale=0.35,angle=270]{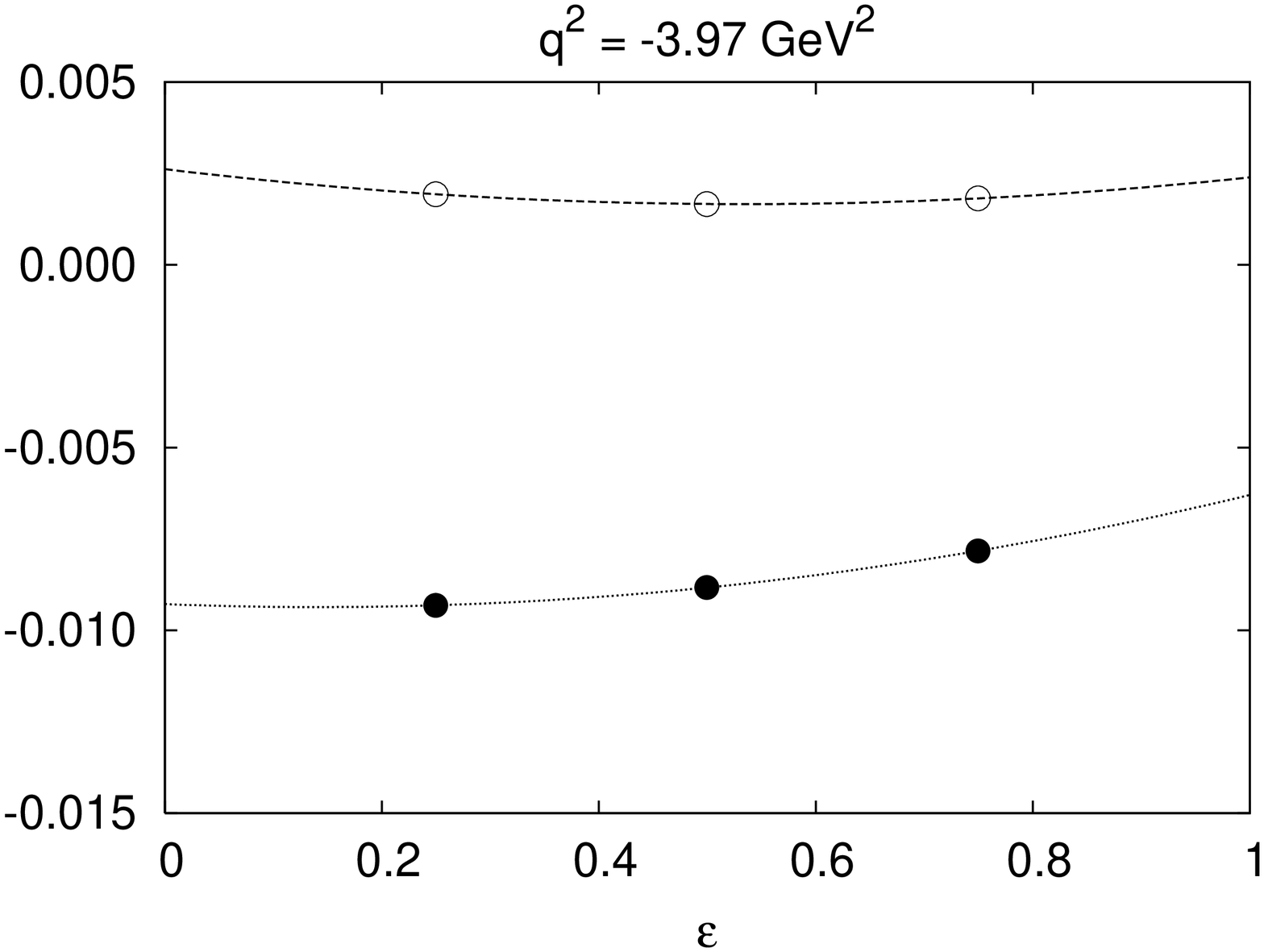}\\
(c)&(d)\\
\hspace{-11mm}\includegraphics[scale=0.35,angle=270]{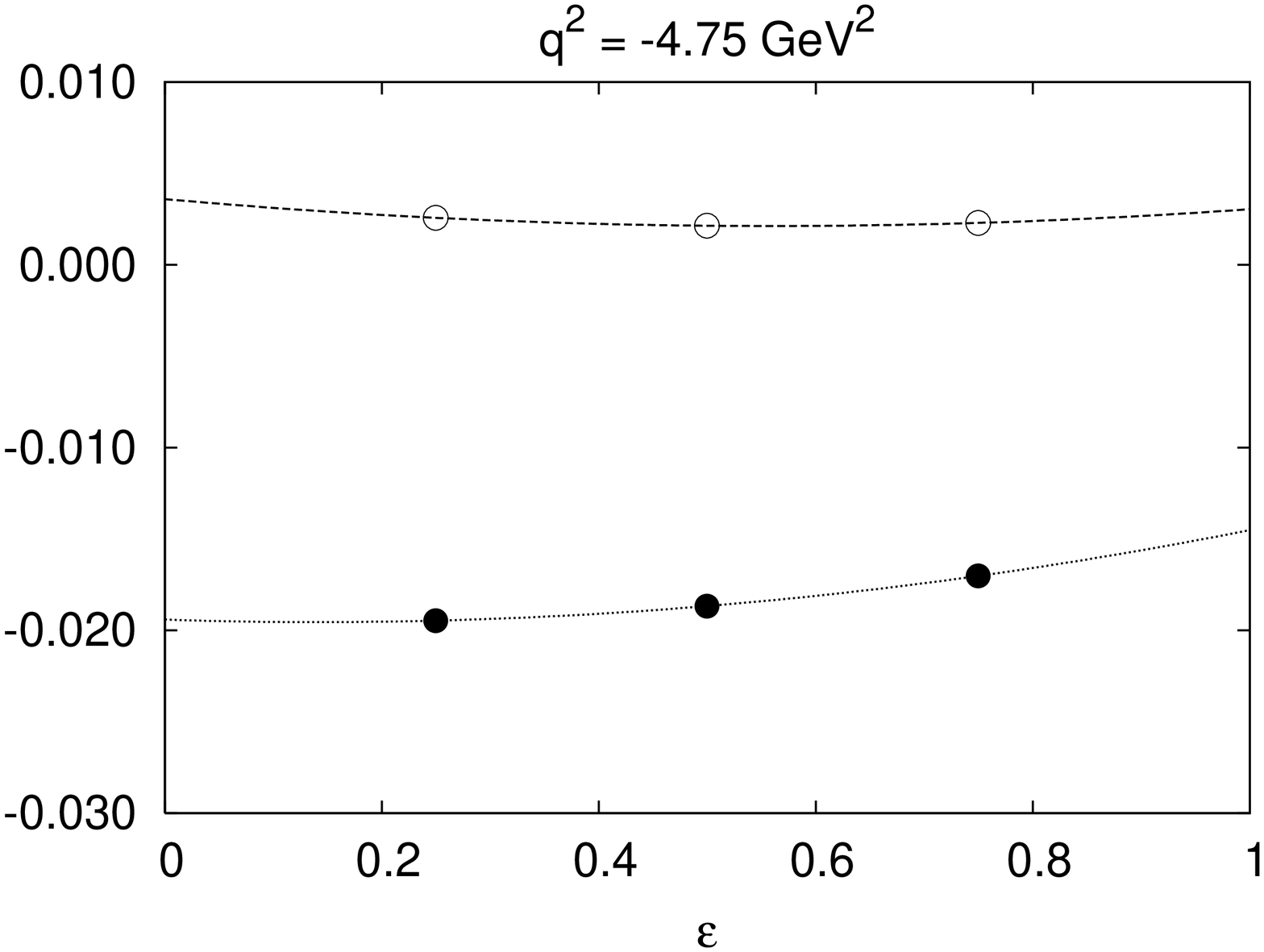}&
\hspace{-11mm}\includegraphics[scale=0.35,angle=270]{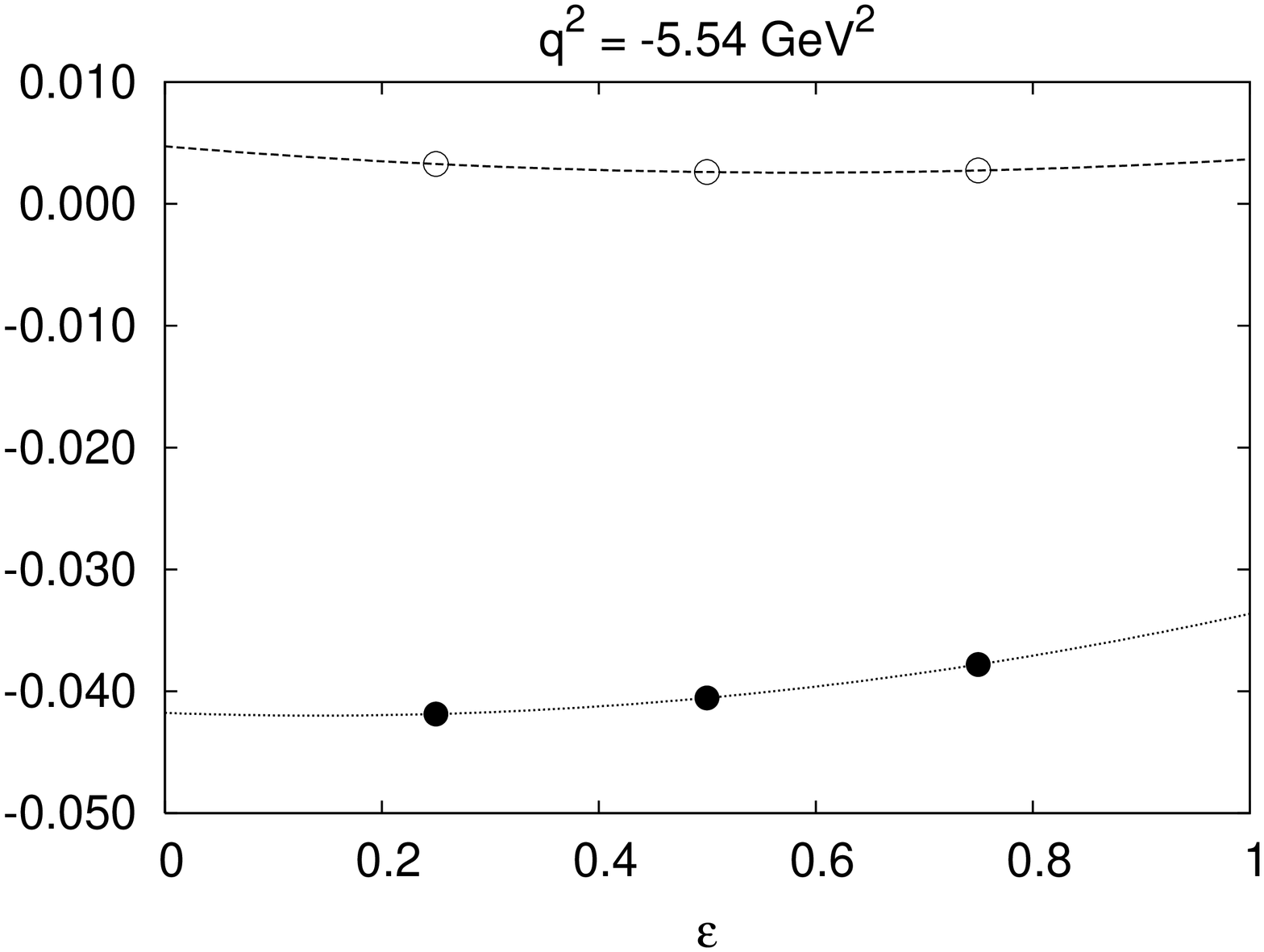}\\
(e)&(f)\\
\end{tabular}
\caption{\small Contribution of the diagrams proportional to $\bar
b^2$  to $\Delta_{\rm L}$ (unfilled circles) and $\Delta_{\rm T}$ (filled
circles) for different $q^2$ - (a) $q^2 = -2.47 GeV^2$, (b) $q^2 =
-2.97 GeV^2$, (c) $q^2 = -3.50 GeV^2$, (d) $q^2 = -3.97 GeV^2$,
(e) $q^2 = -4.75 GeV^2$, (f) $q^2 = -5.54 GeV^2$. $\Delta_{\rm T}$ is
defined in (\ref{eq:delta_t}) and $\Delta_{\rm L}$ is defined in
(\ref{eq:delta_l}). Here $\bar b =1$.} \label{fig:result_bbar}
\end{center}
\end{figure*}

\section{Calculations and Results}
\noindent As explained in \cite{Jain} there are three diagrams
that contribute in the two-photon exchange processes.
Figs.~\ref{fig:twophoton}(a) and (b) show the box and cross-box
diagrams respectively and Fig.~\ref{fig:twophoton}(c) shows a 
diagram proportional to $\bar b^2$.
%\begin{figure*}
%\begin{center}
%\begin{tabular}{cc}
%\hspace{-10mm}\includegraphics[scale=0.35,angle=270]{rto_ep_q2.47.ps}&
%\hspace{-5mm}\includegraphics[scale=0.35,angle=270]{rto_ep_q2.97.ps}\\
%(a)&\hspace{5mm}(b)\\
%\hspace{-10mm}\includegraphics[scale=0.35,angle=270]{rto_ep_q3.50.ps}&
%\hspace{-5mm}\includegraphics[scale=0.35,angle=270]{rto_ep_q3.97.ps}\\
%(c)&\hspace{5mm}(d)\\
%\hspace{-10mm}\includegraphics[scale=0.35,angle=270]{rto_ep_q4.75.ps}&
%\hspace{-5mm}\includegraphics[scale=0.35,angle=270]{rto_ep_q5.54.ps}\\
%(e)&\hspace{5mm}(f)\\
%\end{tabular}
%\caption{\small Dependance of the corrected ratio $\mu_p\bar R$ on
%$\varepsilon$ with $\bar b=0$ (unfilled squares), $\bar b=2$
%(unfilled circles), $\bar b=3$ (unfilled triangles), $\bar b=4$
%(unfilled diamonds), $\bar b=5$ (unfilled pentagon) and $\bar
%b=6.91$ (filled circles).} \label{fig:rto_epsilon}
%\end{center}
%\end{figure*}
The amplitudes for the box and cross-box diagrams are given by:
\ba
\nn i\mathcal{M}_{\rm B} &=& e^4\int\frac{d^4l}{(2\pi)^4}\Bigg[\frac{\bar{u}(k')\gamma^\mu(\slashed{k}-\slashed{l})\gamma^\nu u(k,s_k) }{((k-l)^2 -m_e^2+i\xi)}\Bigg]\\
\nn&&\times\Bigg[\frac{1}{(l^2-\mu^2+i\xi)(\tilde{q}^2-\mu^2 +i\xi)}\Bigg]\\
\nn&&\times\Bigg[\bar{U}(p',s_{p'})\Bigg\{F_1(\tilde{q})\gamma_\mu+ i\frac{\kappa_p}{2M_p}F_2(\tilde{q})\sigma_{\mu\alpha}\tilde{q}^\alpha\Bigg\}\\
\nn&&\times\Bigg\{\frac{\slashed{p}+\slashed{l}+M_p}{(p+l)^2-M_p^2+i\xi}\Bigg\}\\
&&\times  \Bigg\{F_1(l)\gamma_\nu +
i\frac{\kappa_p}{2M_p}F_2(l)\sigma_{\nu\beta}l^\beta\Bigg\}U(p)\Bigg],\\
\nn i\mathcal{M}_{\rm CB} &=& e^4\int\frac{d^4l}{(2\pi)^4}\Bigg[\frac{\bar{u}(k')\gamma^\mu(\slashed{k}-\slashed{l})\gamma^\nu u(k,s_e) }{(k-l)^2 -m_e^2+i\xi}\Bigg]\\
\nn &&\times\Bigg[\frac{1}{(l^2-\mu^2+i\xi)(\tilde{q}^2-\mu^2 +i\xi)}\Bigg]\\
\nn &&\times \Bigg[\bar{U}(p',s_{p'})\Bigg\{F_1(l)\gamma_\nu + i\frac{\kappa_p}{2M_p}F_2(l)\sigma_{\nu\beta}l^\beta\Bigg\}\\
\nn &&\times\Bigg\{\frac{\slashed{p}+\slashed{q}-\slashed{l}+M_p}{(p+\tilde{q})^2-M_p^2+i\xi}\Bigg\}\\
 &&\times\Bigg\{F_1(\tilde{q})\gamma_\mu +
i\frac{\kappa_p}{2M_p}F_2(\tilde{q})\sigma_{\mu\alpha}\tilde{q}^\alpha\Bigg\}U(p)\Bigg].
\ea Here $\tilde{q} = q-l$ and $\xi$ is an infinitesimal
positive parameter. The amplitude for the diagram
proportional to $\bar b^2$ can be written as: \ba
\nn i\mathcal{M}_{\bar b} &=& \Bigg(\frac{e^4\bar b^{2}}{8M_p^2}\Bigg)\int\frac{d^4l}{(2\pi)^4}\Bigg[\frac{\bar{u}(k')\gamma^\mu(\slashed{k}-\slashed{l})\gamma^\nu u(k,s_e) }{(k-l)^2 -m_e^2+i\xi}\Bigg]\\
\nn &&\times \Bigg[\frac{1}{(l^2-\mu^2+i\xi)(\tilde{q}^2-\mu^2 +i\xi)}\Bigg]\\
\nn &&\times \Bigg[\bar{U}(p',s_{p'})\Bigg\{ \Big(\frac{i\kappa_p}{2M_p}F_2(\tilde{q})\sigma_{\mu\alpha}\tilde{q}^\alpha\Big)(\slashed{p}+\slashed{l}-M_p)\\
\nn &&\times \Big(\frac{i\kappa_p}{2M_p}F_2(l)\sigma_{\nu\beta}l^\beta\Big)+ \Big(\frac{i\kappa_p}{2M_p}F_2(l)\sigma_{\nu\beta}l^\beta\Big)\\
\nn &&\times (\slashed{p}+\slashed{q}-\slashed{l}-M_p)
\Big(\frac{i\kappa_p}{2M_p}F_2(\tilde{q})\sigma_{\mu\alpha}\tilde{q}^\alpha\Big)\Bigg\}U(p)\Bigg].\\
\ea
The total two photon exchange amplitude,
\be
\mathcal{M}_{2\gamma} = \mathcal{M}_{\rm B} + \mathcal{M}_{\rm CB} +
\mathcal{M}_{\bar b}\, . \ee
In the numerical calculation we have dropped $m_e$.
The model used for the form factors \cite{Baldini,Baldini_ge} is
the same as Model I described in \cite{Jain}. The values of the
parameters are repeated here in Appendix A for convenience.
The results of our calculation may show some dependence on the precise 
form factor model. This appears to be the main uncertainty in our 
calculation which we hope to explore in a future publication. 

Contributions from box and cross-box diagrams are computed
at 10 different values of $\mu^2$ (from 0.005 to 0.0095)
for each value of $q^2$ and $\varepsilon$. The cross-box
diagram is well defined but for the evaluation of the box
diagram a small imaginary term $\xi$ is kept in the
propagators. This makes the integral in the infrared limit
well defined in the case $m_e=0$. For each value of $q^2$,
$\varepsilon$ and $\mu^2$ we have calculated the box
diagram amplitude for 4 different values of $\xi$ (between
0.001 and 0.00175). The final $\mu^2$ dependent box diagram
amplitudes are obtained by extrapolation to $\xi=0$. The
diagram proportional $\bar b^2$ has no infrared (IR)
divergent term. So the contribution coming from it is
computed keeping $\mu^2 = 0$.

Following Mo and Tsai \cite{Tsai,Mo,Maximon,Ent} one can show that in the
limit $\mu^2 \rightarrow 0$ the leading term from the box and
cross-box diagram can be expressed as: \ba
\mathcal{M}^{2\gamma}_{\rm IR} =
\frac{\alpha}{\pi}[K(p',k)-K(p,k)]\mathcal{M}_0, \ea where \bas
K(p_i,p_j) &=& (p_i.p_j)\int_0^1\frac{dx}{(xp_i+(1-x)p_j)^2}\\
&\times&\ln\left[\frac{(xp_i+(1-x)p_j)^2}{\mu^2}\right]. \eas
\begin{inlinefigure}
\vspace{-5mm}\includegraphics[scale=0.36,angle=270]{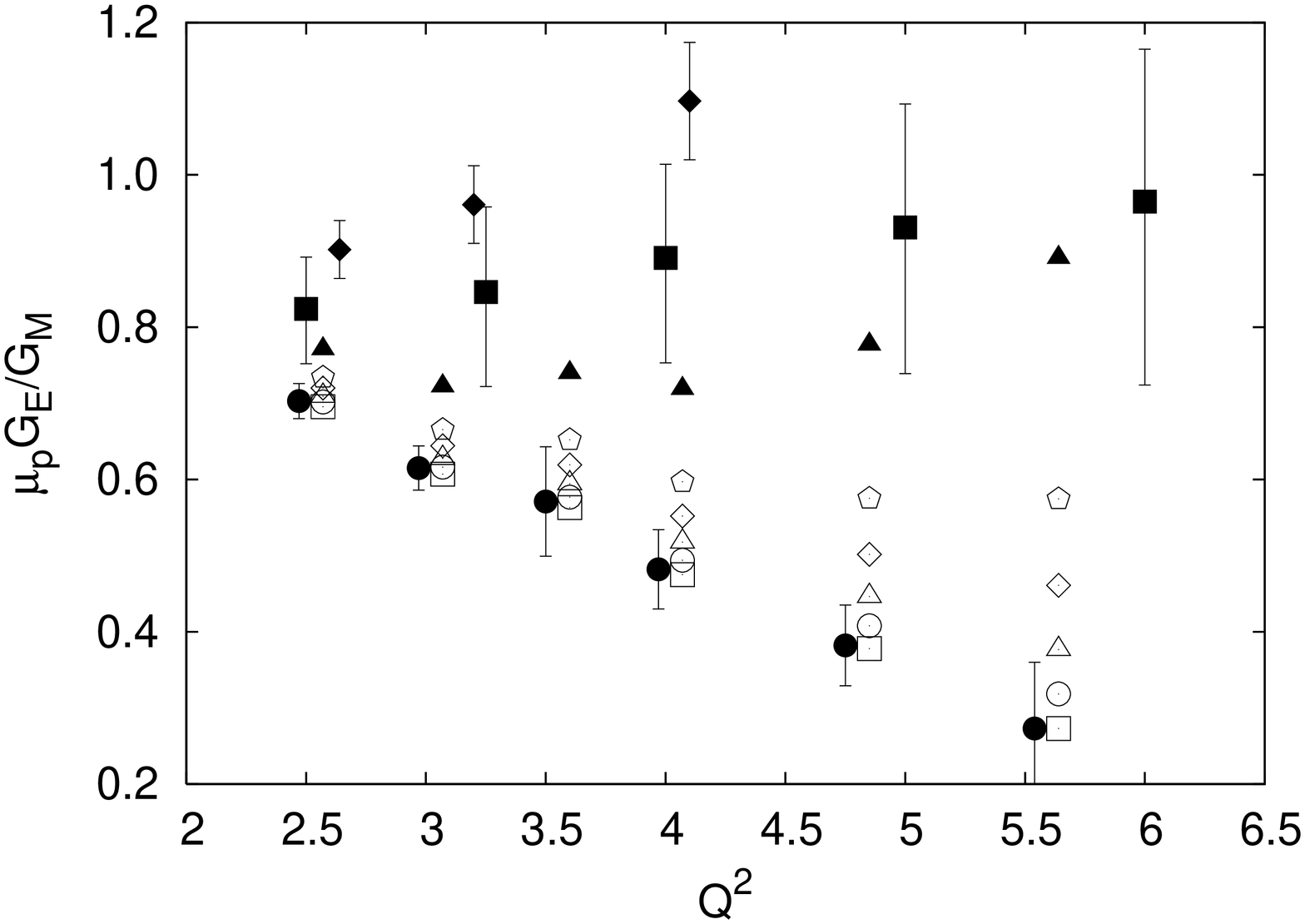}
\caption{\small The ratio $\mu_pG_{\rm E}/G_{\rm M}$ from polarization
transfer experiments (filled circles) and after correction for the
two-photon exchange diagrams with $\bar b=0$ (unfilled squares),
$\bar b=2$ (unfilled circles), $\bar b=3$ (unfilled triangles),
$\bar b=4$ (unfilled diamonds), $\bar b=5$ (unfilled pentagon) and
$\bar b=6.91$ (filled triangles). The corrected values have been
offset by $\Delta Q^2 = 0.1\, GeV^2$ for clarity. The filled
diamonds and the filled squares represent the Rosenbluth
extraction experiment results obtained at JLAB and SLAC
respectively. } \label{fig:ratio}
\end{inlinefigure}
\begin{inlinefigure}
\vspace{-5mm}\includegraphics[scale=0.36,angle=270]{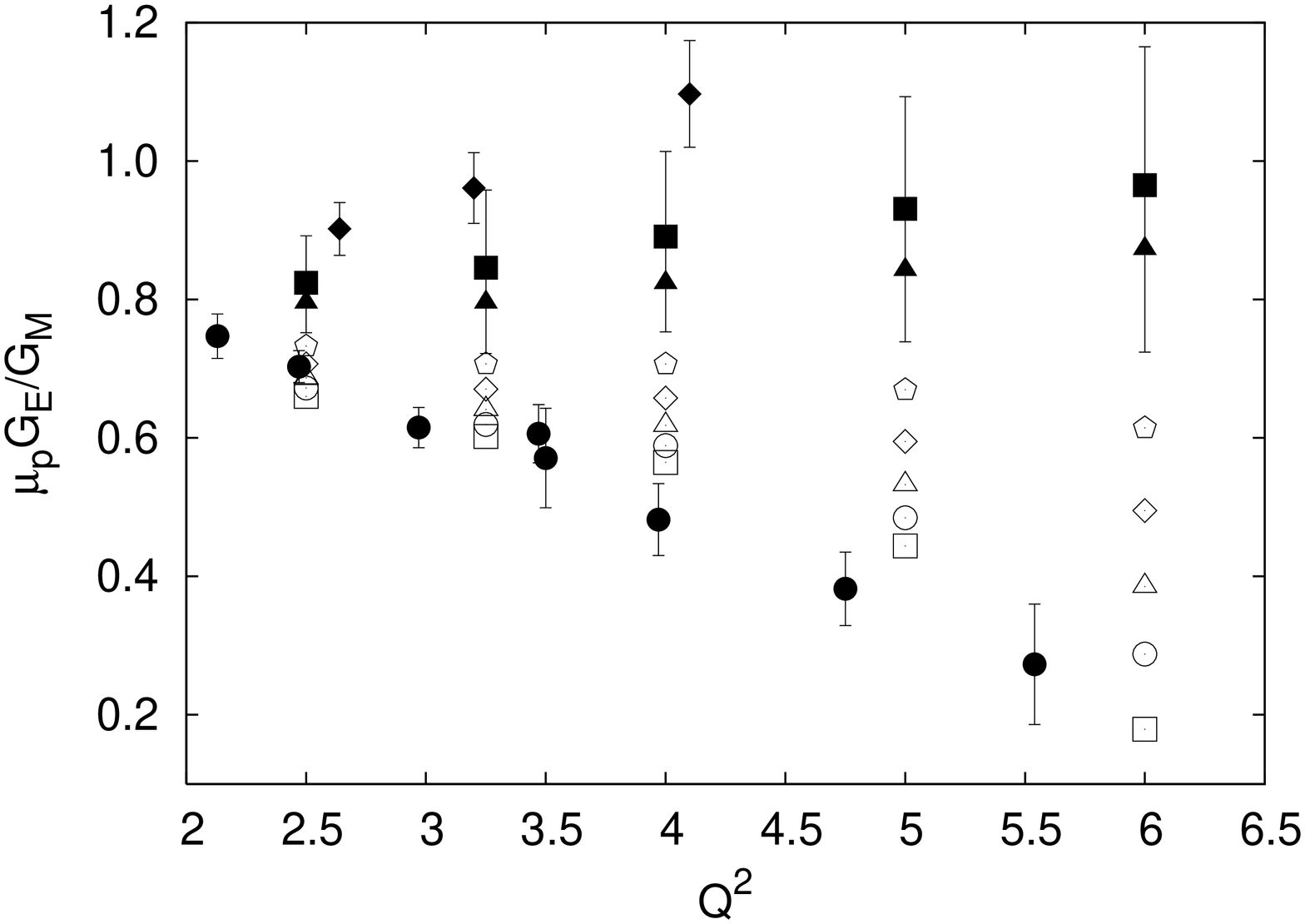}
\caption{\small The ratio $\mu_pG_{\rm E}/G_{\rm M}$ from SLAC Rosenbluth
extraction experiments (filled squares) and after correction for
the (unpolarized) two-photon exchange diagrams with $\bar b=0$
(unfilled squares), $\bar b=2$ (unfilled circles), $\bar b=3$
(unfilled triangles), $\bar b=4$ (unfilled diamonds), $\bar b=5$
(unfilled pentagon) and $\bar b=6.91$ (filled triangles). The
filled diamonds represent the JLAB Rosenbluth extraction
experiment results and the filled circles represent the results of
the polarization transfer experiments. } \label{fig:ratio_unpol}
\end{inlinefigure}
\vspace{1mm}

\noindent Hence, the leading contributions to the cross sections
coming from the box and cross-box diagrams are given by: \ba
\left.\frac{d(\Delta\sigma^{2\gamma}_{\rm L,T})}{d\Omega_e}\right|_{\rm IR}
=
\frac{2\alpha}{\pi}[K(p',k)-K(p,k)]\frac{d(\Delta\sigma^{1\gamma}_{\rm L,T})}{d\Omega_e}
. \ea Let \ba
\left.\frac{d(\Delta\sigma^{2\gamma}_{\rm L,T})}{d\Omega_e}\right|_{\rm IR}\equiv
a^{ir}_{\rm L,T} + b^{ir}_{\rm L,T}\ln\mu^2. \ea
\begin{figure*}
\begin{center}
\begin{tabular}{cc}
\hspace{-11mm}\includegraphics[scale=0.35,angle=270]{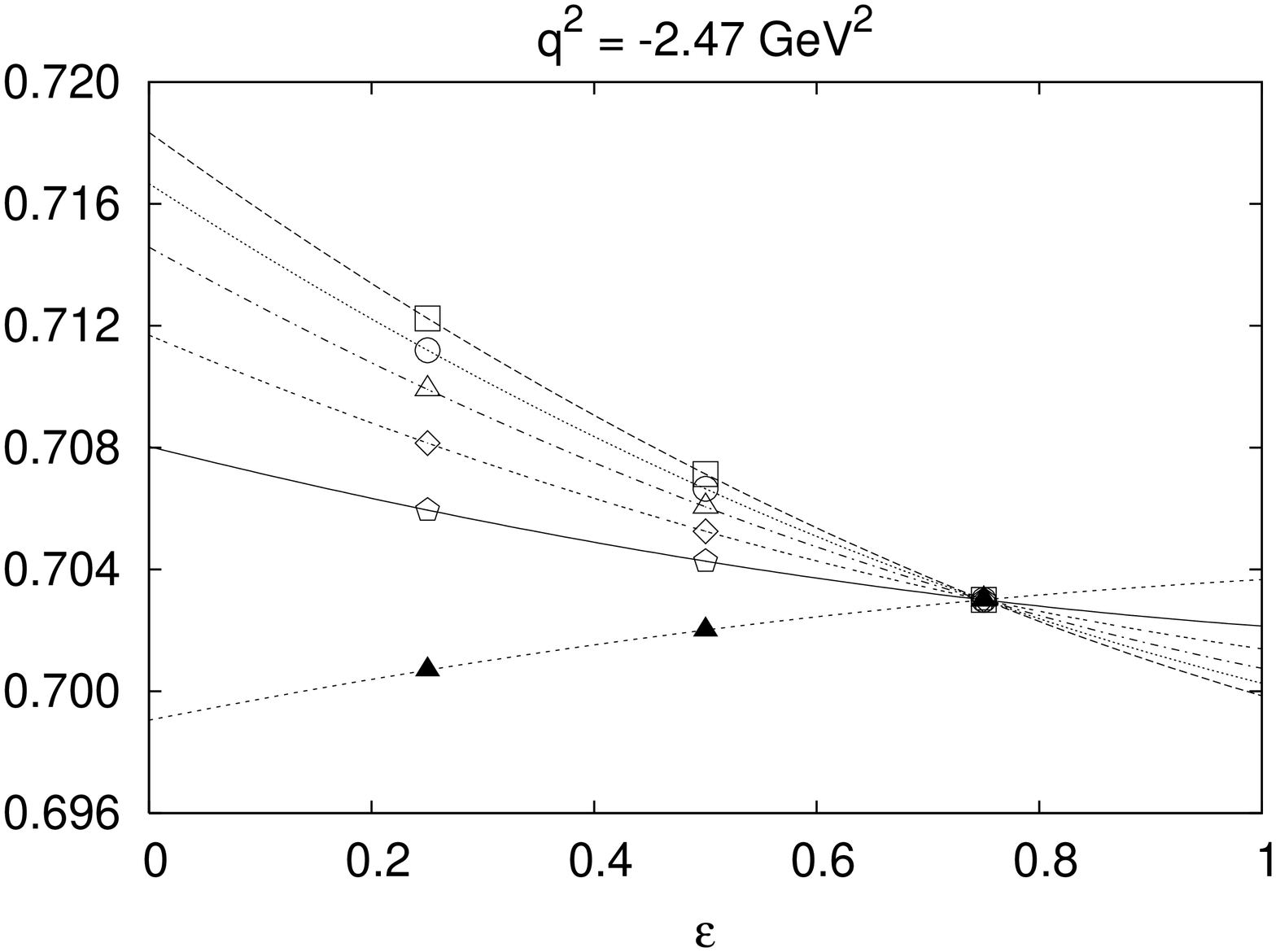}&
\hspace{-11mm}\includegraphics[scale=0.35,angle=270]{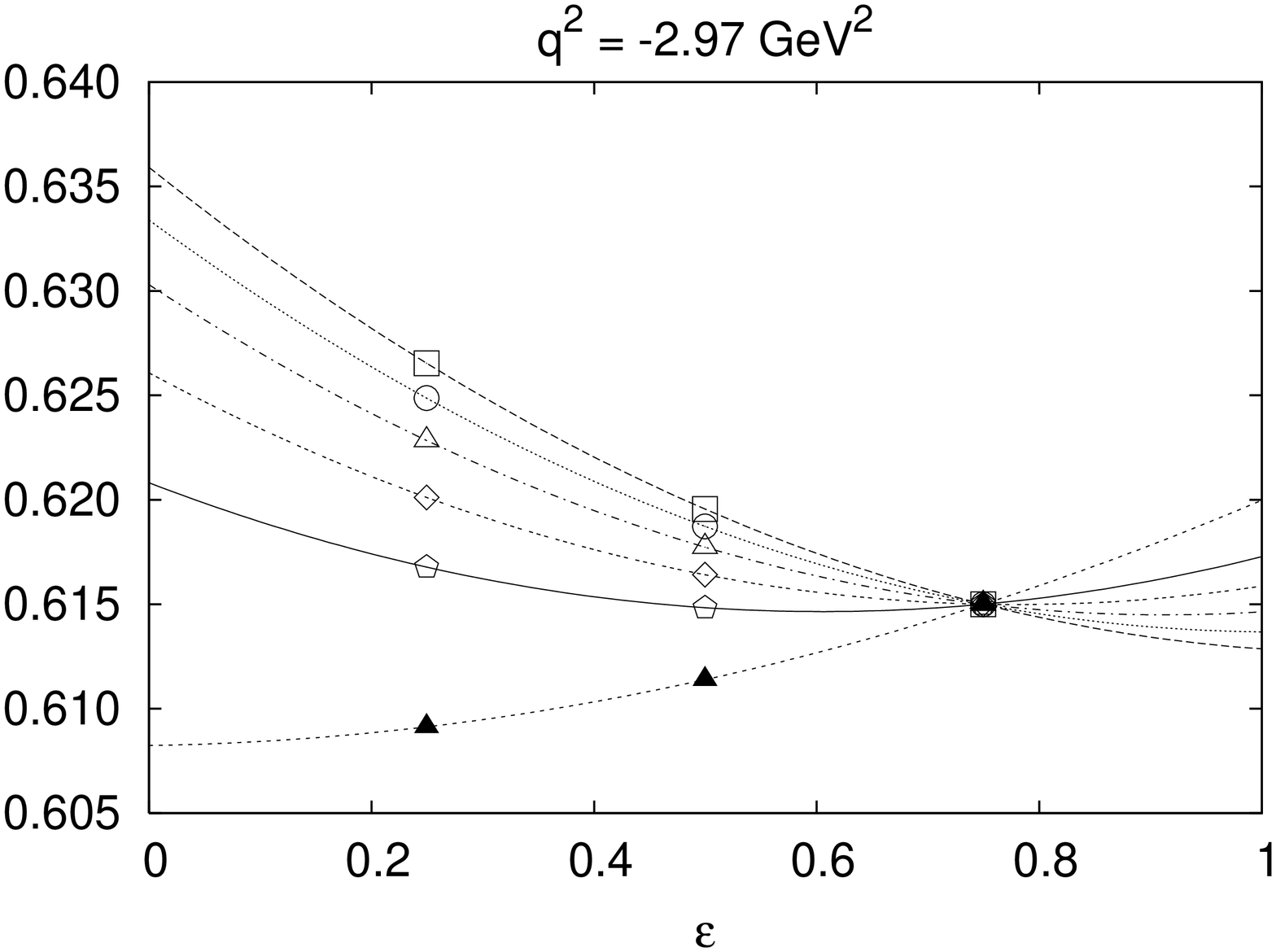}\\
(a)&(b)\\
\hspace{-11mm}\includegraphics[scale=0.35,angle=270]{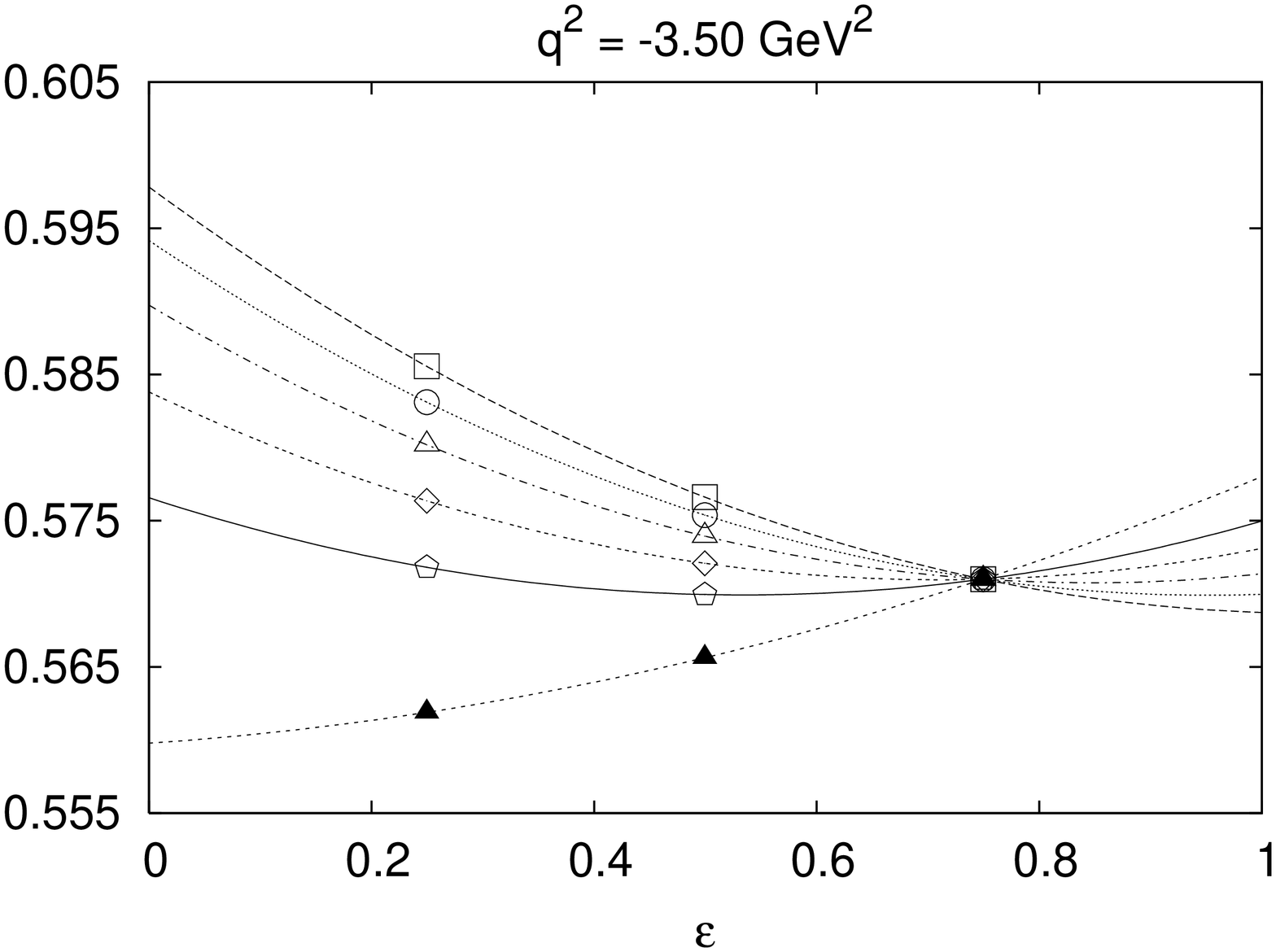}&
\hspace{-11mm}\includegraphics[scale=0.35,angle=270]{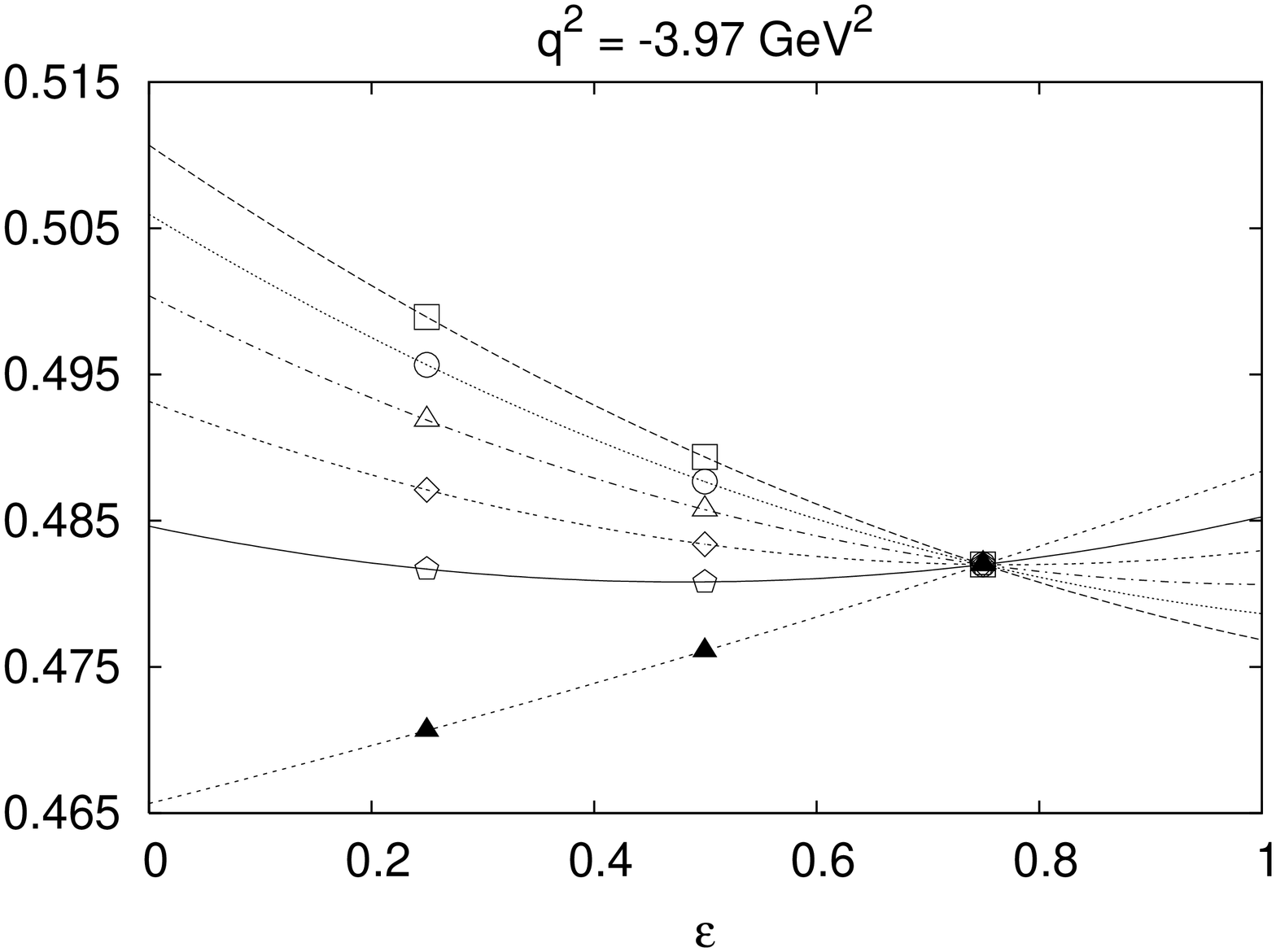}\\
(c)&(d)\\
\hspace{-11mm}\includegraphics[scale=0.35,angle=270]{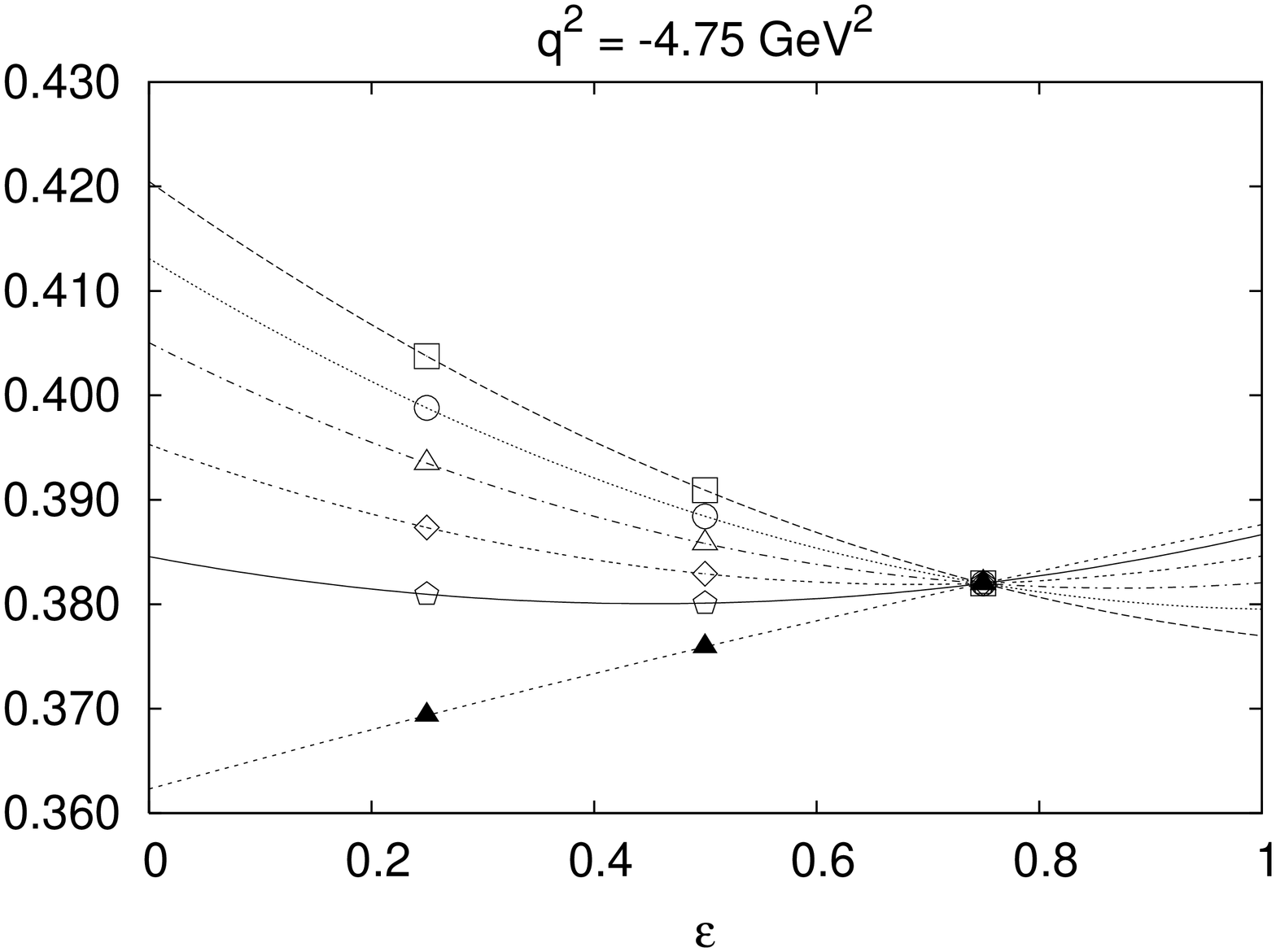}&
\hspace{-11mm}\includegraphics[scale=0.35,angle=270]{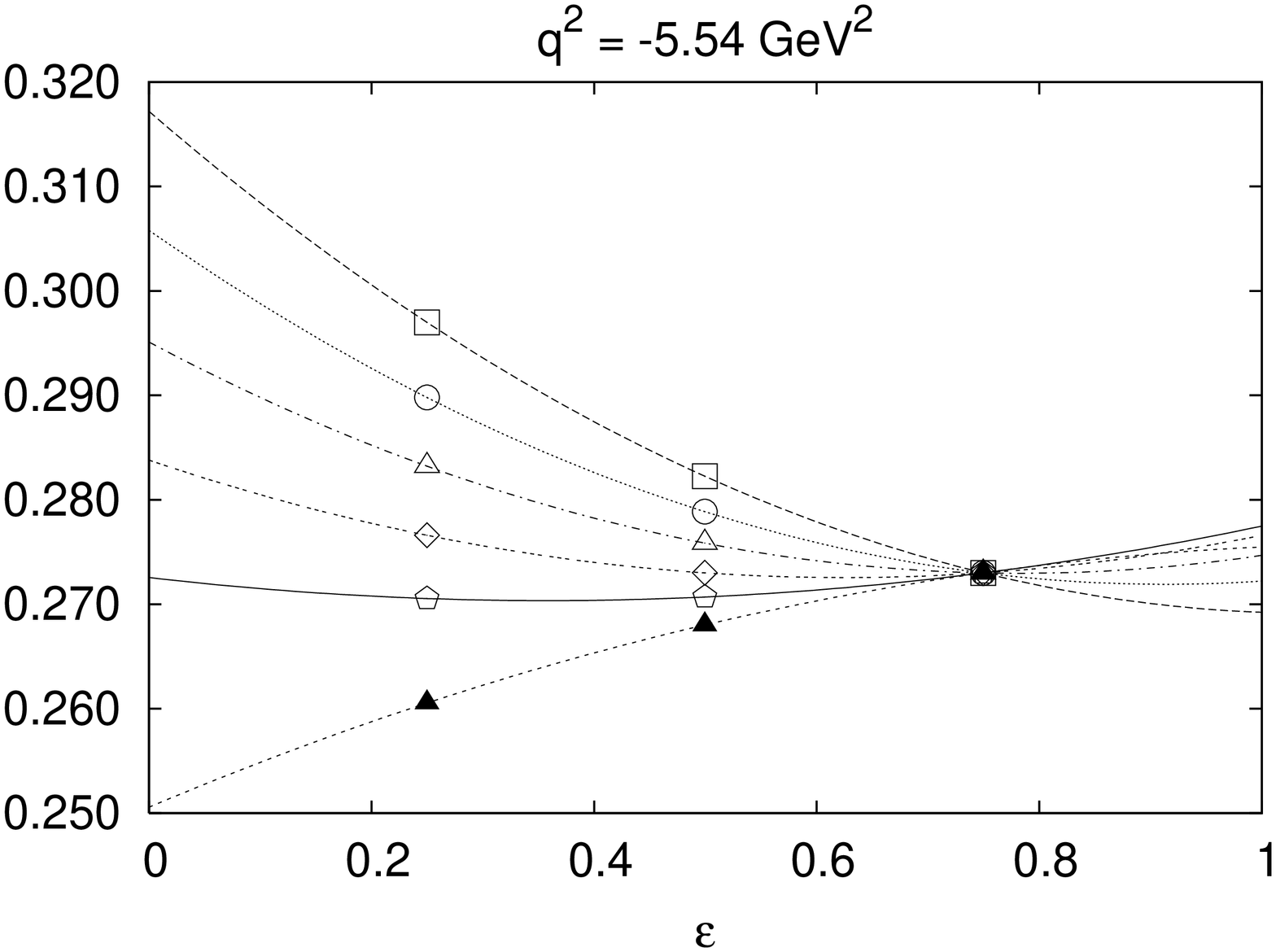}\\
(e)&(f)\\
\end{tabular}
\caption{\small Dependance of the form factor ratio $\mu_p G_{\rm E}/G_{\rm M}$ on $\varepsilon$ from different $q^2$ - (a) $q^2 = -2.47 GeV^2$, (b) $q^2 =
-2.97 GeV^2$, (c) $q^2 = -3.50 GeV^2$, (d) $q^2 = -3.97 GeV^2$,
(e) $q^2 = -4.75 GeV^2$, (f) $q^2 = -5.54 GeV^2$. For each $q^2$, the ratio, $\mu G_{\rm E}/G_{\rm M}$ is plotted for different $\bar b$ values - $\bar b=0$ (unfilled squares),
$\bar b=2$ (unfilled circles), $\bar b=3$ (unfilled triangles),
$\bar b=4$ (unfilled diamonds), $\bar b=5$ (unfilled pentagon) and
$\bar b=6.91$ (filled triangles).} \label{fig:prediction}
\end{center}
\end{figure*}
To remove the IR part from
$d(\Delta\sigma^{2\gamma}_{\rm L,T})/d\Omega_e$ we fit these with the
following functions: \ba
\frac{d(\Delta\sigma^{2\gamma}_{\rm L,T})}{d\Omega_e} = a_{\rm L,T}(\mu^2) +
b_{\rm L,T}(\mu^2)\ln\mu^2 \ea with \bas
a_{\rm L,T}(\mu^2) &=& a^0_{\rm L,T} + a^1_{\rm L,T}\,\mu^2 + \mathcal{O}[\mu^4]\\
b_{\rm L,T}(\mu^2) &=& b^{ir}_{\rm L,T} + b^1_{\rm L,T}\,\mu^2 +
\mathcal{O}[\mu^4]. \eas Hence, $a^0_{\rm L,T}$ give the IR
removed $d(\Delta\sigma^{2\gamma}_{\rm \rm L,T})/d\Omega_e$.

Fig. \ref{fig:result_bcb} shows the contribution to $\Delta_{\rm L,T}$
coming from box and cross-box diagrams for different $q^2$. The
contribution from the diagram proportional to $\bar b^2$ is shown
in Fig. \ref{fig:result_bbar}.  %Fig. \ref{fig:rto_epsilon} shows the dependance of $\mu_p\bar R$ on
%$\varepsilon$ for different $\bar b$.
Fig. \ref{fig:ratio} shows the ratio $\mu_pG_{\rm E}/G_{\rm M}$ from
polarization transfer experiments (filled circles) and after
correction for the two-photon exchange diagrams with $\bar b=0$
(unfilled squares), $\bar b=2$ (unfilled circles), $\bar b=3$
(unfilled triangles), $\bar b=4$ (unfilled diamonds), $\bar b=5$
(unfilled pentagon) and $\bar b=6.91$ (filled triangles). The
highest value of $\bar b$ used is obtained from the limit on
nonlinearity of JLAB Rosenbluth data (see Section 3). %The corrected ratio $\mu_p\bar R$ shows some non-linear dependance on $\varepsilon$.
In the range $3.5\ge Q^2\ge 2.5$ GeV$^2$, 
the polarization transfer experiment 
used the kinematic regime with
$\varepsilon$ lying roughly between 0.7 and 0.8 \cite{Punjabi}.  
For simplicity, here we assume
a fixed value $\varepsilon=0.75$.
Fig. \ref{fig:ratio_unpol} shows
the effect of $\bar b$ on the unpolarized SLAC data. The method
used to obtain these corrections is described in Section 6 of
\cite{Jain}.

The form factor ratio $G_{\rm E}/G_{\rm M}$ depends only on $Q^2$. However
the ratio $R_{\rm E}$, defined in (\ref{eq:RE}),
may show a dependence on $\varepsilon$ due to higher order contributions.
We predict this dependence by assuming that the electron-proton elastic
scattering cross section can be well approximated by including only one and
two photon exchange diagrams. The uncorrected
form factor ratio $R_{\rm E}$ can be calculated by using (\ref{eq:Rbar}),
where the corrected ratio $\bar R$ is independent of $\varepsilon$.
Fig. \ref{fig:prediction} shows the predicted dependance of the uncorrected
form factor ratio $\mu_p R_{\rm E}$ on $\varepsilon$ for different $\bar b$ and $q^2$. Here again we have assumed that the kinematic regime of the polarization
transfer experiment corresponds to a fixed $\varepsilon=0.75$.
This may be used in future to fix the value of the parameter $\bar b$.

\section{Limit on $\bar b$}
As described in \cite{Jain} $\bar b$ is a free parameter. But the fact
that $\sigma^{\bar b}_{\rm R}$ varies nonlinearly with $\varepsilon$
allows us to put a limit on $\bar b$. We fit the reduced
cross-section for the scattering of unpolarized electrons from
protons, $\sigma_{\rm R}$ with the following function: \ba \sigma_{\rm R} =
\mc P_0 + \mc P_1\bar\varepsilon + \mc P_2\bar
\varepsilon^2,\label{chisq_fit} \ea where $\bar \varepsilon =
\varepsilon - \frac12$. As the result obtained by JLAB
\cite{Qattan,Qattan2} contains much smaller error-bars than the SLAC data
we use JLAB data to obtain $\mc P_2$ for different $Q^2=-q^2$.
Then we fit the contribution of the box and cross-box diagrams
$(\sigma^{\rm BCB}_{\rm R})$ and the contribution proportional to $\bar b^2$
($\sigma^{\bar b}_{\rm R}$) to the reduced cross-section with similar
functions, i.e., \ba
\sigma^{\rm BCB}_{\rm R} &=& \mc R^{\rm BCB}_0 + \mc R^{\rm BCB}_1\bar\varepsilon + \mc R^{\rm BCB}_2\bar \varepsilon^2,\label{fit_bcb}\\
\sigma^{\bar b}_{\rm R} &=& \bar b^2(\mc R^{\bar b}_0 + \mc R^{\bar b}_1\bar\varepsilon + \mc R^{\bar b}_2\bar \varepsilon^2).\label{fit_bbar}
\ea

To put the limit on $\bar b$ we first obtain a limit on the nonlinearity
parameter $\mc P_2$. The 2 sigma limit is obtained by finding the largest
value of $-\mc P_2$ such that $\chi^2$ deviates from its minimum value
by 4 units. We consider the largest value of $-\mc P_2$ since the two photon
exchange contributions are found to give a negative value of $\mc P_2$.
The parameters $\mc P_0$, $\mc P_1$
are also allowed to vary while determining the maximum value allowed for
$-\mc P_2$. The best fit value of $\mc P_2$ and the 1 and 2 sigma limits
are given in Table \ref{tab1}. The best fit values of $\mc R^{\rm BCB}_2$
and $\mc R^{\bar b}_2$ are given in Table \ref{tab2}.
 Assuming that the dominant source of nonlinear behaviour are
the box, cross-box and $\bar b$ diagrams, we obtain the limit on
$\bar b^2$ by the following relation, \ba \bar b^2_{max} =
\frac{\mc P_{2limit} - \mc R^{\rm BCB}_2}{\mc R^{\bar b}_2} \ea where
$\mc P_{2limit}$ is the one or two sigma limit on the parameter
$\mc P_2$. From Table \ref{tab1} we see that the most stringent
limit on $\bar b$ is obtained for $Q^2 = 2.64$ GeV$^2$ and is
given by: $|\bar b| \lesssim 6.91$.
The nonlinearity of the reduced cross section $\sigma_{\rm R}$ has also been computed
in \cite{Carlson} using the GPD formalism.
\bigskip

%\vspace{5mm}
\begin{inlinetable}
\begin{tabular}{ccccc}
\hline\\
$Q^2$ & $\mc P_{2}$ & $\mc P_{2}$ & $\mc P_{2}$   & $\bar b^2$ \\
in    & best       & 1 sigma    &  2 sigma  &  2 sigma\\
GeV$^2$ & fit  &  limit &  limit & limit \\\\
\hline\hline\\
2.64&$2.02$&$-3.82$&$-9.64$&$47.74$\\
\\
3.20&$1.10$&$-3.70$&-8.52&$70.61$\\
\\
4.10&$2.68$&$-2.92$&-8.48&$152.93$\\\\
\hline
\end{tabular}
\caption{\small The best fit and the one and two sigma limit on
the parameter $\mc P_2$ (defined
in (\ref{chisq_fit})) for different $Q^2$. All values of
$\mc P_2$ have been scaled by $10^4$. The corresponding two sigma limit
on the parameter $\bar b^2$ is also given.}
\label{tab1}
\end{inlinetable}

\vspace{5mm}
\begin{inlinetable}
%\begin{table*}
\begin{tabular}{ccc}
\hline\\
$Q^2$ in GeV$^2$&$\mc R^{\rm BCB}_{2}$ &$\mc R^{\bar b}_2$\\\\
\hline\hline\\
2.64&$-4.04\times 10^{-4}$&$-11.73\times 10^{-6}$\\
\\
3.20&$-3.69\times 10^{-4}$&$-6.84\times 10^{-6}$\\
\\
4.10&$-1.95\times 10^{-4}$&$-4.27\times 10^{-6}$\\\\
\hline
\end{tabular}
\caption{\small Values of the fitting parameters $\mc R^{\rm BCB}_2$
(defined in (\ref{fit_bcb})) and  $\mc R^{\bar b}_2$ (defined in
(\ref{fit_bbar})) for different $Q^2$. } \label{tab2}
\end{inlinetable}
%\end{table*}

%\vspace{5mm}
\section{Conclusion}
We have computed the two photon exchange corrections to the proton
electromagnetic form factor ratio $\mu_p G_{\rm E}/G_{\rm M}$ using a gauge
invariant nonlocal Lagrangian. The Lagrangian is truncated to
dimension five operators and depends on one unknown parameter
$\bar b$. The higher dimension operators are expected to give
negligible contributions as long as the off-shellness of the
proton propagator is small. The two photon exchange corrections to
the ratio are found to be small as long as the parameter $\bar
b\sim 1$. We impose a limit on this parameter by the predicted
nonlinearity in the $\varepsilon $ dependence of the unpolarized
reduced cross section $\sigma_{\rm R}$ due to two photon exchange
contributions. A two sigma limit is found to be $|\bar b|< 6.91$.
For such large value of $\bar b$ the corrections to the
polarization transfer experiment are quite significant. The
corrections are found to go in the right direction and make the
polarization transfer results come very close to the SLAC
Rosenbluth measurement of this ratio. 
Hence we find that for a wide range of values of $\bar b$
the results of the two experiments agree within errors after including
the two photon exchange contributions.
We also predict an $\varepsilon$ dependence of the
ratio extracted from polarization transfer. This can be tested in
future experiments and also be used to extract the value of the
parameter $\bar b$.

\appendix
\renewcommand{\thesection}{Appendix \Alph{section}}
\renewcommand{\theequation}{\Alph{section}.\arabic{equation}}
\setcounter{equation}{0}

\section{Model for the Form Factors}

\noindent The fits for $G_{\rm M}/\mu_p$ and $G_{\rm E}$ are given by: \ba
\frac{G_{\rm M}(q^2)}{\mu_p} =
\sum_{a=1}^4\frac{A^{'}_a}{(q^2-m_a^2+im_a\Gamma^{'}_a)}\label{gmbymu}\\
G_{\rm E}(q^2) =
\sum_{a=1}^6\frac{B^{'}_a}{(q^2-m_a^2+im_a\Gamma^{'}_a)}\label{ge}.
\ea The values of the masses and the parameters are tabulated in
Table \ref{tab3}. Using the fits for the magnetic and electric
form factors one can determine the Dirac and Pauli form factors.

\vspace{5mm}
\begin{inlinetable}
\begin{tabular}{lrrrr}
\hline\\
$a$&$A^{'}_a~~~$&$B^{'}_a~~~$&$m_a~~$&$\Gamma^{'}_a~~$\\\\
\hline\hline\\
1&$-2.882564$&$-3.177877$&0.8084&0.2226\\
&$+i\, 1.944314$&$+i\,2.123389$&&\\
2&$2.882564 $&$3.177877$&0.9116&0.1974\\
&$- i\, 1.944314$&$ - i\,2.123389$&&\\
3&$-1.064011$&$-0.608148$&1.274&0.5712\\
&$ - i\,3.216318$&$ - i\,5.685885$&&\\
4&$1.064011$&$0.608148$&1.326&0.5488\\
&$+ i\,3.216318$&$ + i\,5.685885$&&\\
5&0~~~~~~~&$3.211388$&1.96&1.02\\
&&$+ i\,0.693412$&&\\
6&0~~~~~~~&$- 3.211388$&2.04&0.98\\
&&$ - i\,0.693412$&&\\
\hline
\end{tabular}
\caption{\small Masses, widths and parameter values for
$G_{\rm M}/\mu_p$ and $G_{\rm E}$ fits. $A^{'}$s and $B^{'}$s are
defined in (\ref{gmbymu}) and (\ref{ge}).} \label{tab3}
\end{inlinetable}

\end{multicols}
\end{document}